\documentclass[twoside,twocolumn,9pt]{article}
\usepackage{extsizes}
\usepackage[super,sort&compress,comma]{natbib} 
\usepackage[version=3]{mhchem}
\usepackage[left=1.5cm, right=1.5cm, top=1.785cm, bottom=2.0cm]{geometry}
\usepackage{balance}
\usepackage{widetext}
\usepackage{times,mathptmx}
\usepackage{sectsty}
\usepackage{graphicx} 
\usepackage{lastpage}
\usepackage[format=plain,justification=raggedright,singlelinecheck=false,font={stretch=1.125,small,sf},labelfont=bf,labelsep=space]{caption}
\usepackage{float}
\usepackage{fancyhdr}
\usepackage{fnpos}
\usepackage[english]{babel}
\usepackage{array}
\usepackage{droidsans}
\usepackage{charter}
\usepackage[T1]{fontenc}
\usepackage[usenames,dvipsnames]{xcolor}
\usepackage{setspace}
\usepackage[compact]{titlesec}

\usepackage{graphics}
\usepackage{amssymb}
\usepackage{pstricks}
\usepackage{rotating,subfigure}
\usepackage{gensymb}
\usepackage{amsmath}

\definecolor{cream}{RGB}{222,217,201}

\begin{document}

\pagestyle{fancy}
\thispagestyle{plain}
\fancypagestyle{plain}{

\fancyhead[C]{\includegraphics[width=18.5cm]{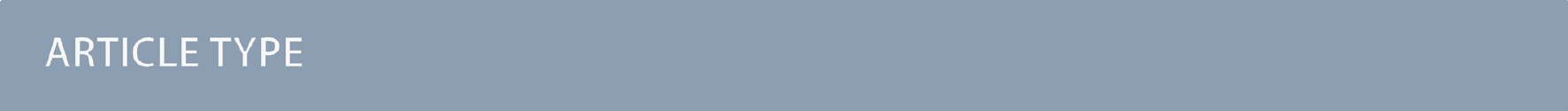}}
\fancyhead[L]{\hspace{0cm}\vspace{1.5cm}\includegraphics[height=30pt]{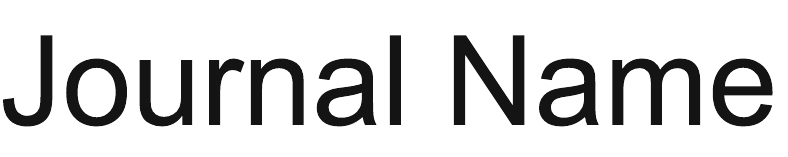}}
\fancyhead[R]{\hspace{0cm}\vspace{1.7cm}\includegraphics[height=55pt]{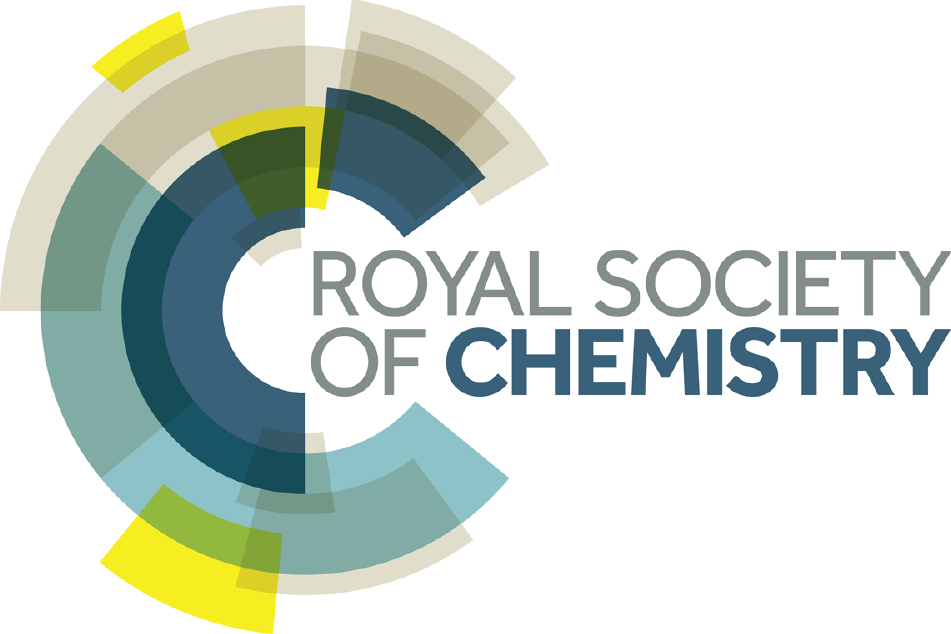}}
\renewcommand{\headrulewidth}{0pt}
}

\makeFNbottom
\makeatletter
\renewcommand\LARGE{\@setfontsize\LARGE{15pt}{17}}
\renewcommand\Large{\@setfontsize\Large{12pt}{14}}
\renewcommand\large{\@setfontsize\large{10pt}{12}}
\renewcommand\footnotesize{\@setfontsize\footnotesize{7pt}{10}}
\makeatother

\renewcommand{\thefootnote}{\fnsymbol{footnote}}
\renewcommand\footnoterule{\vspace*{1pt}%
\color{cream}\hrule width 3.5in height 0.4pt \color{black}\vspace*{5pt}} 
\setcounter{secnumdepth}{5}

\makeatletter 
\renewcommand\@biblabel[1]{#1}            
\renewcommand\@makefntext[1]%
{\noindent\makebox[0pt][r]{\@thefnmark\,}#1}
\makeatother 
\renewcommand{\figurename}{\small{Fig.}~}
\sectionfont{\sffamily\Large}
\subsectionfont{\normalsize}
\subsubsectionfont{\bf}
\setstretch{1.125} 
\setlength{\skip\footins}{0.8cm}
\setlength{\footnotesep}{0.25cm}
\setlength{\jot}{10pt}
\titlespacing*{\section}{0pt}{4pt}{4pt}
\titlespacing*{\subsection}{0pt}{15pt}{1pt}

\fancyfoot{}
\fancyfoot[LO,RE]{\vspace{-7.1pt}\includegraphics[height=9pt]{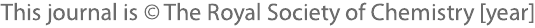}}
\fancyfoot[CO]{\vspace{-7.1pt}\hspace{13.2cm}\includegraphics{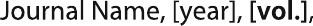}}
\fancyfoot[CE]{\vspace{-7.2pt}\hspace{-14.2cm}\includegraphics{RF}}
\fancyfoot[RO]{\footnotesize{\sffamily{1--\pageref{LastPage} ~\textbar  \hspace{2pt}\thepage}}}
\fancyfoot[LE]{\footnotesize{\sffamily{\thepage~\textbar\hspace{3.45cm} 1--\pageref{LastPage}}}}
\fancyhead{}
\renewcommand{\headrulewidth}{0pt} 
\renewcommand{\footrulewidth}{0pt}
\setlength{\arrayrulewidth}{1pt}
\setlength{\columnsep}{6.5mm}
\setlength\bibsep{1pt}

\makeatletter 
\newlength{\figrulesep} 
\setlength{\figrulesep}{0.5\textfloatsep} 

\newcommand{\topfigrule}{\vspace*{-1pt}%
\noindent{\color{cream}\rule[-\figrulesep]{\columnwidth}{1.5pt}} }

\newcommand{\botfigrule}{\vspace*{-2pt}%
\noindent{\color{cream}\rule[\figrulesep]{\columnwidth}{1.5pt}} }

\newcommand{\dblfigrule}{\vspace*{-1pt}%
\noindent{\color{cream}\rule[-\figrulesep]{\textwidth}{1.5pt}} }

\makeatother

\renewcommand{\log}{\mathrm{log}}
\renewcommand{\vec}{\mathbf}
\newcommand{\nwater}{N_{\mathrm{H}_2\mathrm{O}}}

\newcommand{\CL}[1]{{\color{magenta}{CL #1 CL}}}
\newcommand{\mike}[1]{{\color{red}{ #1}}}

\twocolumn[
  \begin{@twocolumnfalse}
\vspace{3cm}
\sffamily
\begin{tabular}{m{4.5cm} p{13.5cm} }

\includegraphics{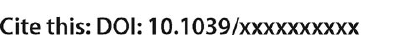} & \noindent\LARGE{\textbf{Cooee bitumen: Dynamics and structure of
  bitumen-water mixtures}} \\
\vspace{0.3cm} & \vspace{0.3cm} \\

 & \noindent\large{Claire A. Lemarchand,$^{\ast}$\textit{$^{a}$} Michael L. Greenfield,\textit{$^{b}$} and Jesper S. Hansen\textit{$^{a}$}} \\

\includegraphics{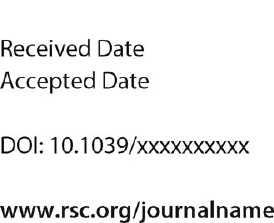} & \noindent\normalsize{Systems of Cooee bitumen and water up to $4$ mass \% are studied by molecular dynamics
simulations. The cohesive energy density of the system is shown to decrease with an increasing water content.
This decrease is due mainly to an increase in potential energy which is not high
enough to counterbalance the increase in volume due to the addition of water.
It is not due to a decrease of potential energy between the slightly polar
asphaltene molecules.
The water molecules tend to form a  droplet in bitumen. The size and the distribution of
sizes of the droplets are quantified, with multiple droplets being more
  stable at the highest temperature simulated. The droplet is mainly located close
to the saturates molecules in Cooee bitumen.
Finally, it is shown that the water dynamics is much slower in bitumen than in pure water
because it is governed by the diffusion of the droplet and not of the single molecules.

} \\

\end{tabular}

 \end{@twocolumnfalse} \vspace{0.6cm}

  ]

\renewcommand*\rmdefault{bch}\normalfont\upshape
\rmfamily
\section*{}
\vspace{-1cm}


\footnotetext{\textit{$^{a}$~DNRF Centre ``Glass and Time'', IMFUFA,\\
  Department of Sciences, Roskilde University,
  Universitetsvej 1, Postbox 260,
  DK-4000 Roskilde, Denmark. E-mail: clemarch@rip.ens-cachan.fr}}
\footnotetext{\textit{$^{b}$~Department of Chemical Engineering,
  University of Rhode Island,
  Kingston, Rhode Island USA 02881 }}





\section{Introduction}
Water is one of the main causes for pavement deterioration.~\cite{airey, cui, blackman}
The mechanism by which water causes pavement failure is complicated and multiple:
water may create a loss of adhesion between the bitumen binder and
the aggregates, and/or a loss of cohesion inside
the bitumen, and/or the bitumen-filler mastic.~\cite{airey, cui, blackman}
On the other hand, water can be used beneficially in the context of road pavements. Bitumen emulsions,
which are formed by injecting a few percent of water into hot bitumen
are used to pave roads at much
moderate temperatures, between $13$\degree C and $23$\degree C.~\cite{muthen}
The emulsion has a much lower viscosity than the bitumen itself allowing for
an easier mixing with the aggregates.~\cite{muthen}
After the mixing step, the water evaporates over several months
leaving bitumen with $0$ to $4$\%
in mass of water in its driest state.~\cite{muthen}

One of the reasons why the role of water in bitumen is so
complex and rich is that bitumen is very hydrophobic,~\cite{salou, kasongo}
while some of the large
aromatic molecules in bitumen, known as asphaltene molecules, are slightly polar.
They are believed to stabilize the water-in-oil emulsion~\cite{mullins2011}
and even assemble more closely in the presence of water.~\cite{gray}
This complex interplay between water and bitumen 
explains why the mechanism by which water causes pavement deterioration is still debated.
In this context, the aim of this article is to address two main questions:
(i) How is the structure and internal cohesion of bitumen changed in the
presence of water? (ii) How and how fast do the water molecules travel inside the bitumen?

To answer these two questions, we use molecular dynamics (MD) simulations 
of a modified model bitumen, known as Cooee bitumen.~\cite{hansen_jcp_2013}
This model bitumen contains four molecule types chosen to resemble the
Hubbard-Stanfield classification.~\cite{hubbard:1948} The asphaltene and resin molecules
of this model contain $2$ and $1$ sulfur atoms, respectively. 
The sulfur atom is chosen because it is by far the most
common heteroatom in most bitumens.~\cite{shrp645}
The polarity
due to the presence of this heteroatom is taken into account in
the MD simulations by adding 
a partial charge on the sulfur atom and on the neighboring carbon atoms.
Several systems with a growing concentration of water molecules are considered.
For each of them, the cohesion and structure of bitumen are quantified using
the cohesive energy of each molecule type, the radial distribution function and
the average size of the nanoaggregates in bitumen.
The structure and dynamics of the water molecules are quantified
using the Delaunay tessellation, mean squared displacement
and hydrogen bond dynamics. 

The paper is organized as follows. Section~\ref{sec:simuDetails}
contains the necessary information about the molecular model and
the simulations. In Sec.~\ref{sec:results}, results on the bitumen
cohesive energy and structure as well as on the structure and dynamics of water
are discussed.
Section~\ref{sec:conclu} contains a summary and a conclusion.

\section{Molecular model and simulation details}
\label{sec:simuDetails}
\begin{figure}
  \begin{center}
  \scalebox{0.23}{\includegraphics[clip=true, trim={8cm 0 0 0}]{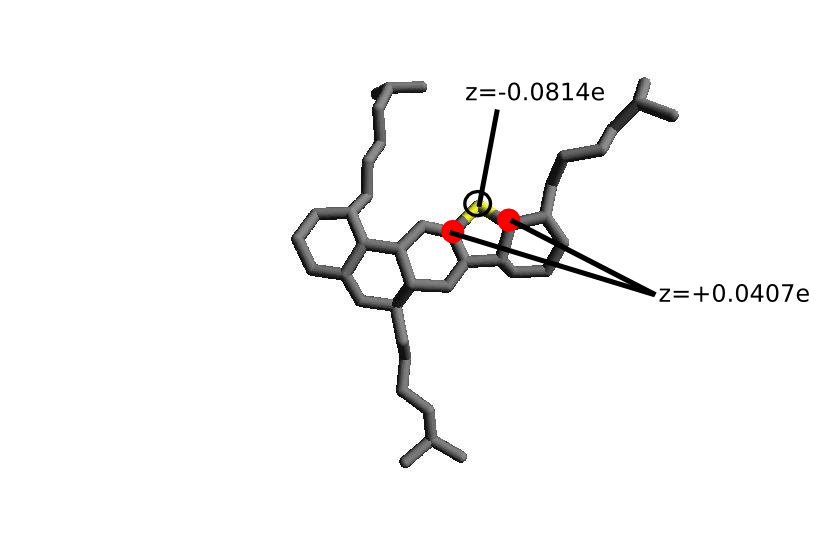}}
  \scalebox{0.13}{\includegraphics{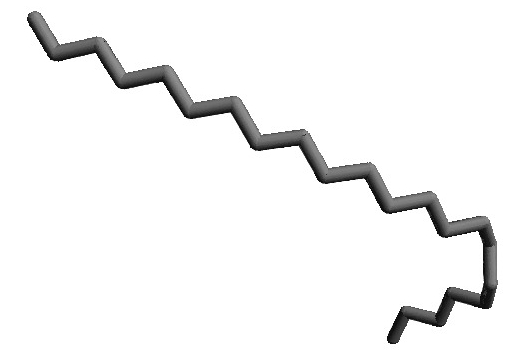}} \\
  \scalebox{0.18}{\includegraphics{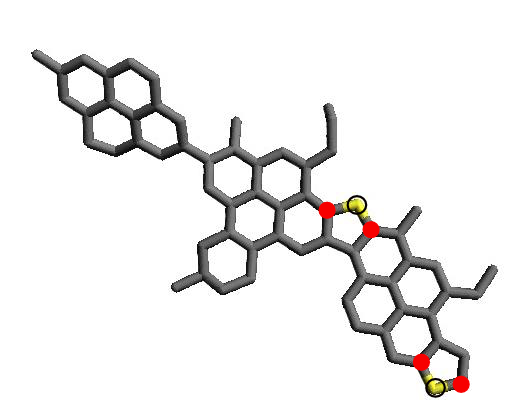}}
  \scalebox{0.18}{\includegraphics{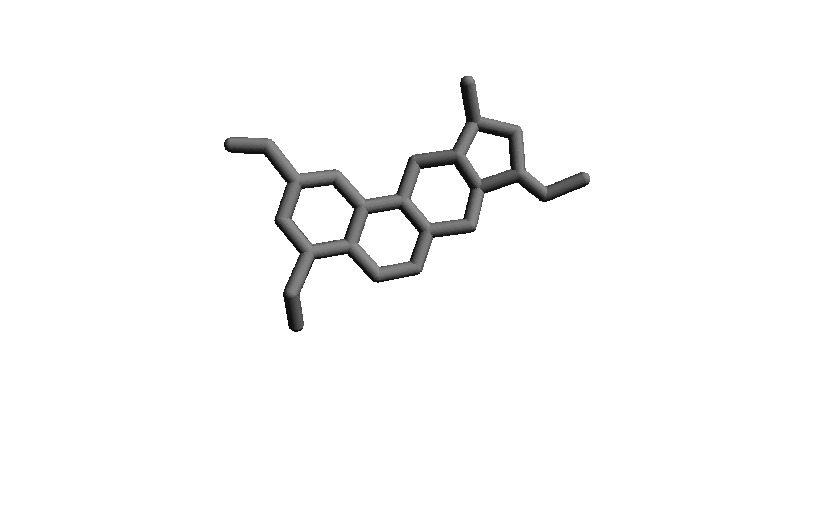}}
  \caption{\label{fig:molstructure} (Color online). Constituent
    molecules in the Cooee bitumen model.  Top left: Resin. Top right:
    Docosane (saturated hydrocarbon).  Lower left: Asphaltene. Lower right:
    Resinous oil.  Yellow indicates sulfur atoms having partial charge
    $z=0.0814 e$, and red circles indicate methylene group with
    partial charge $z=0.0407 e$, these are labeled $C_2$.  Hydrocarbon
    groups with zero charge are labelled $C_1$.  $e$ is the
    fundamental unit of charge $e=1.602 \times 10^{-19}$ C.}
  \end{center}
\end{figure}

The molecular model is the Cooee bitumen model~\cite{hansen_jcp_2013}
which consists of four constituent molecule types, representing the
asphaltene, resin, resinous oil and hydrocarbon components in bitumen.
The structures of the four molecules chosen are presented in Fig. \ref{fig:molstructure}. 
This classification is based on the Hubbard-Stanfield scheme.~\cite{hubbard:1948,rostler:1965}
In the original model, methyl,
methylene, and methine groups are represented by the same
Lennard-Jones particle, \textit{i.e.}, a united atomic unit (UAU). The sulfur
atom is also a Lennard-Jones particle, but with a different mass. In
the presence of water, electrostatic interactions may be important, and
in this work we include these into the Cooee model using a simple
point charge model. In this way, the two methylene groups forming
bonds with a sulfur are labeled $C_2$ and are each given a partial
positive charge of $z=0.0407 e$ \cite{liu:2010} as this group is less
electronegative than sulfur. Likewise the sulfur is given a
partial negative charge of $z=-0.0817 e$, ensuring charge
neutrality. Carbon-based groups not forming bonds with sulfur are
labeled $C_1$ and have zero charge. 

The force field is a simple extension of the original Cooee bitumen
force field:
\begin{equation}
\begin{aligned}
  U(\vec{r})&=   \sum_{i} \sum_{j>i} 4\epsilon_{ij}\left[
          \left(\frac{\sigma_{ij}}{r_{ij}}\right)^{12} -
          \left(\frac{\sigma_{ij}}{r_{ij}}\right)^{6}
          \right] + \sum_{i} \sum_{j>i}
          \frac{z_iz_j}{4\pi\epsilon_0r_{ij}} + \\
            &\mbox{}
            \frac{1}{2}\sum_{\text{\tiny{bonds}}}k_s(r_{ij}-l_{\text{\tiny{b}}})^2 + 
            \frac{1}{2}\sum_{\text{\tiny{angles}}}k_{\theta}(\cos \theta - \cos
            \theta_0)^2 + 
  \sum_{\text{\tiny{dihedrals}}}\sum_{n=0}^5 c_n\cos^n \phi \ .
  \label{eq:forcefield}
\end{aligned}
\end{equation}
The first two sums deal with pair interactions and the last three with the
intramolecular interactions. Parameters for the intramolecular
interactions in bitumen are given in Ref.~\cite{hansen_jcp_2013} The
pair-interaction parameters are listed in Table
\ref{table:paircoeff} in SI units. The simulations are performed
in reduced units such that $\sigma_{C_1C_1} = 1$, $\epsilon_{C_1C_1} = 1$,
$m_{C_1} = 1$, $k_B = 1$ and $4\pi\epsilon_0 = 1$.
The parameters between particles with different
Lennard-Jones interactions are given by the Lorentz-Berthelot mixing
rules.~\cite{AllenTildesley} To reduce the computational time, we approximate the
electrostatic force calculations by a shifted force method, \textit{i.e.}, the
force is 
\begin{equation}
F_c = \frac{z_iz_j}{4\pi \epsilon_0}\left(\frac{1}{r_{ij}^2} -
  \frac{1}{r_c^2}\right) \ \ \mbox{if} \ \ r_{ij} \leq r_c \ ,
\end{equation}
using a cutoff of $r_c=16.9$ \AA. For non-confined systems the shifted-force
approximation performs surprisingly well
\cite{hansen_2012,takahashi_2011} and is applicable here. 

The water model is based on the SPC/Fw model
\cite{toukan_1985,wu_2006} which is a flexible three-site
model. The force field is given by Eq. (1), with intramolecular
parameters $k_s/k_B = 268 089$ K \AA$^{-2}$, $k_\theta/k_B=38 152$ K rad$^{-2}$,
$l_b = 1.012$ \AA, and $\theta_0=1.91$ rad.
Due to computational efficiency, 
the values of these parameters are modified compared to Refs.~\cite{wu_2006, raabe_2007}
such that the equilibrium angle value $\theta_0$ is closer to the experimental value
and the bonds are less rigid.
The pair interaction parameters are listed in Table
\ref{table:paircoeff}. 
\begin{table}[h]
\small
  \caption{\ Pair interaction parameters for the bitumen and water system.
  Parameters between particles with different
  Lennard-Jones interactions are given by the Lorentz-Berthelot mixing
  rules}
  \label{table:paircoeff}
  \begin{tabular*}{0.5\textwidth}{@{\extracolsep{\fill}}lllll}
    \hline
     $X$  & $\sigma_{XX}$ (\AA) & $\epsilon_{XX}/k_B$ (T) & $m$ (g/mol) & $z$ ($e$)  \\
    \hline
C$_1$ &  3.75          &  75.4             & 13.3         & 0        \\
C$_2$ &  3.75          &  75.4             & 13.3         & 0.0407   \\
S     &  3.75          &  75.4             & 32.0         & -0.0814  \\
O     &  3.15          &  78.4             & 16.0         & 0.82     \\
H     &   -            &  0.0              & 1.0          & -0.41    \\
    \hline
  \end{tabular*}
\end{table}

In all simulations we use $30$ resin, $30$ resinous oil, $30$ asphaltene
and $246$ docosane molecules.  Different systems
with varying water contents are investigated, namely, with $n_W =
0, 5, 10, 20, 40, 70, 100, 150, 200$ and $300$ water molecules.
It corresponds to a mass fraction of water varying from $0$ to $4$\%.
The mass density is chosen so that the average pressure is equal to $1$ atm. 
Three temperatures are investigated in detail, namely, $T = 603.2$ K,
$T = 452.4$ K, and $T = 377.0$ K 
(reduced temperatures $T^*=k_BT/\epsilon_{C_1C_1} = 8.0, 6.0, 5.0$).
For these three temperatures and for all systems,
eight independent initial configurations are considered.
The equilibration time for each simulation is of $50$ million steps
followed by a production run of $50$ million time steps also.
It corresponds to $43$ ns.
Moreover, for all systems, one simulation run
is performed for temperatures ranging from $T=603.2$ K to $T = 301.6$ K
by steps of $37.7$ K. It enables us to see the effect of temperature
with more accuracy.
For these simulations the equilibration period is $40$ million time steps
and the production period $20$ million time steps.
These extensive runs are performed on a GeForce GTX 780 Ti graphics card using
the RUMD software package~\cite{rumd} version 3.0.

The sizes of water droplets were determined using a geometric
analysis.~\cite{greenfield93} The simulation box was subdivided into
tetrahedra using Delaunay tessellation. Atoms are located at each vertex, and
the Delaunay algorithm~\cite{tanemura83} ensures that no atoms are located within a
tetrahedron.  Droplets were defined by (1) identifying tetrahedra with at least
two vertices being oxygen or hydrogen atoms of a water molecule, (2)
noting which of these tetrahedra shared common faces,
and (3) iteratively grouping together such tetrahedra with the neighbors
  of their neighboring tetrahedra until all regions that share connectivity
  were connected. These are called ``clusters'' in the language of
Ref.~\cite{greenfield93} Periodic boundary conditions were accounted for
when defining each tetrahedron. Restricting the tetrahedra to those with 3
or 4 atoms being from water leads to similar results as those shown below.

The volume of a droplet reported here equals the sum of the volumes of
tetrahedra in a cluster.  The water molecules within a droplet correspond to
the water molecules that define the tetrahedra in a cluster.  The number of
droplets was determined by choosing a minimum cluster volume of 1~\AA$^3$.

Geometry calculations were performed on configurations taken each
$2^{20}=1.05\times 10^6$ time steps (48 per production run) for 4 of the 8
independent runs.  Results were averaged separately for each combination of
droplets, i.e.\ separate averages and distributions were calculated for
drops 1 to $n$ (in decreasing volume) 
when $n$ droplets happened to be present.  These results were combined to
create averages over the largest droplet in each configuration.

\section{Results and discussion}
\label{sec:results}

\subsection{Bitumen cohesion and structure}
\label{sec:bitumen}

The presence of water is believed to reduce the cohesion
inside bitumen.~\cite{airey, cui, blackman} The MD simulations allow us to check this hypothesis
and also to study more precisely which molecule types are the
most affected in their internal cohesion and structure in the presence of water.

\begin{figure*}
  \begin{center}
  \subfigure[]{\includegraphics[scale=0.25]{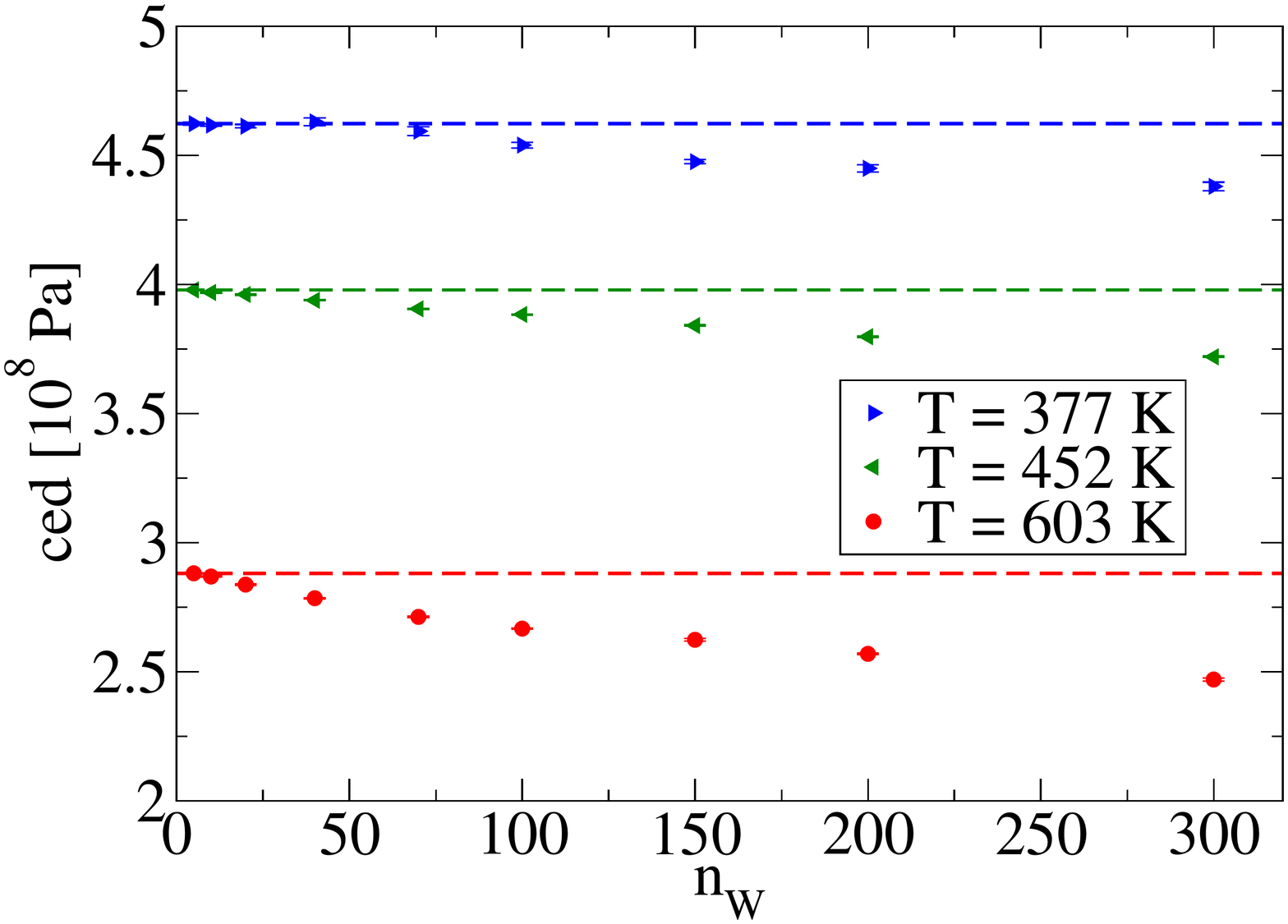}}
  \subfigure[]{\includegraphics[scale=0.25]{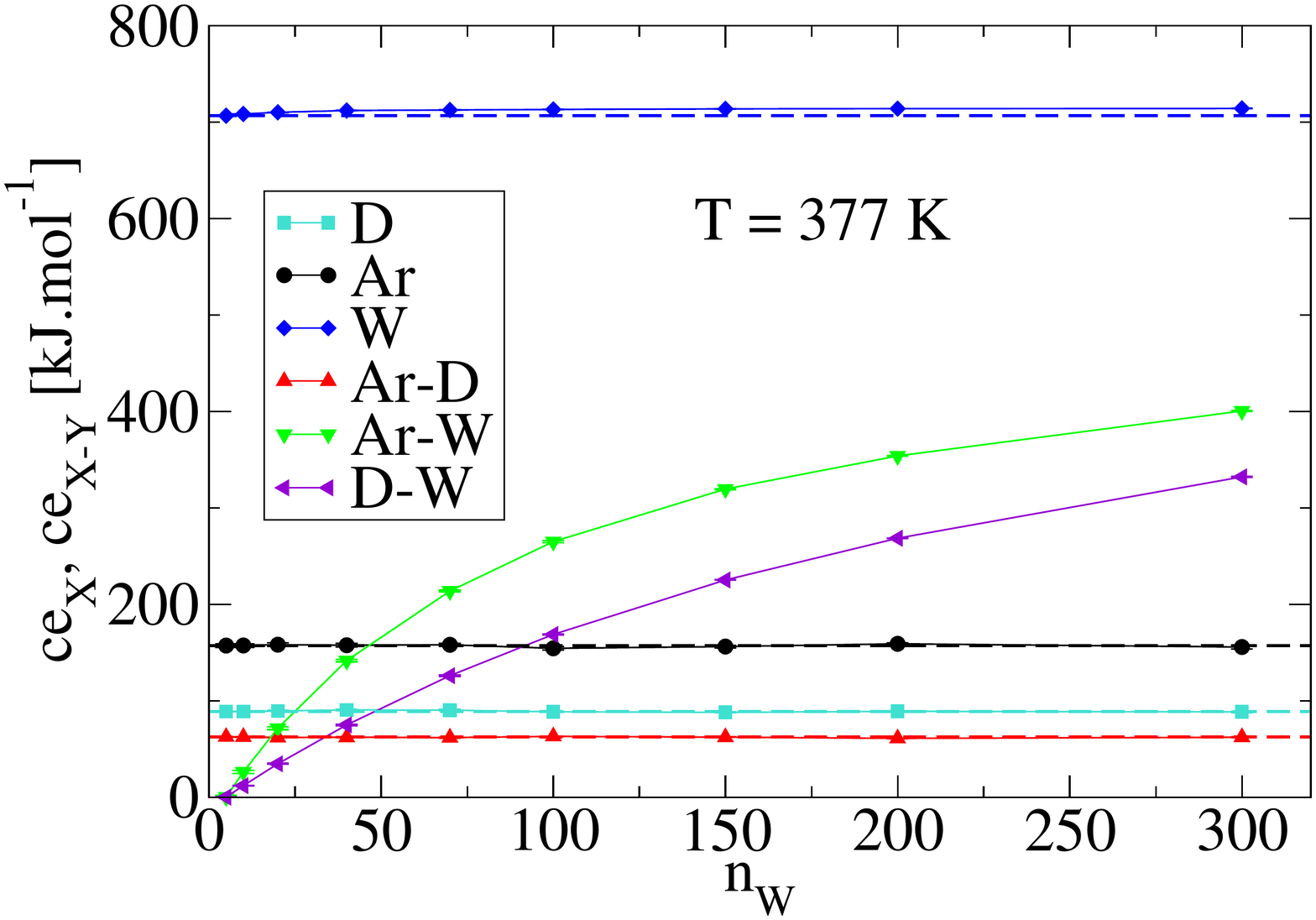}}\\
  \subfigure[]{\includegraphics[scale=0.25]{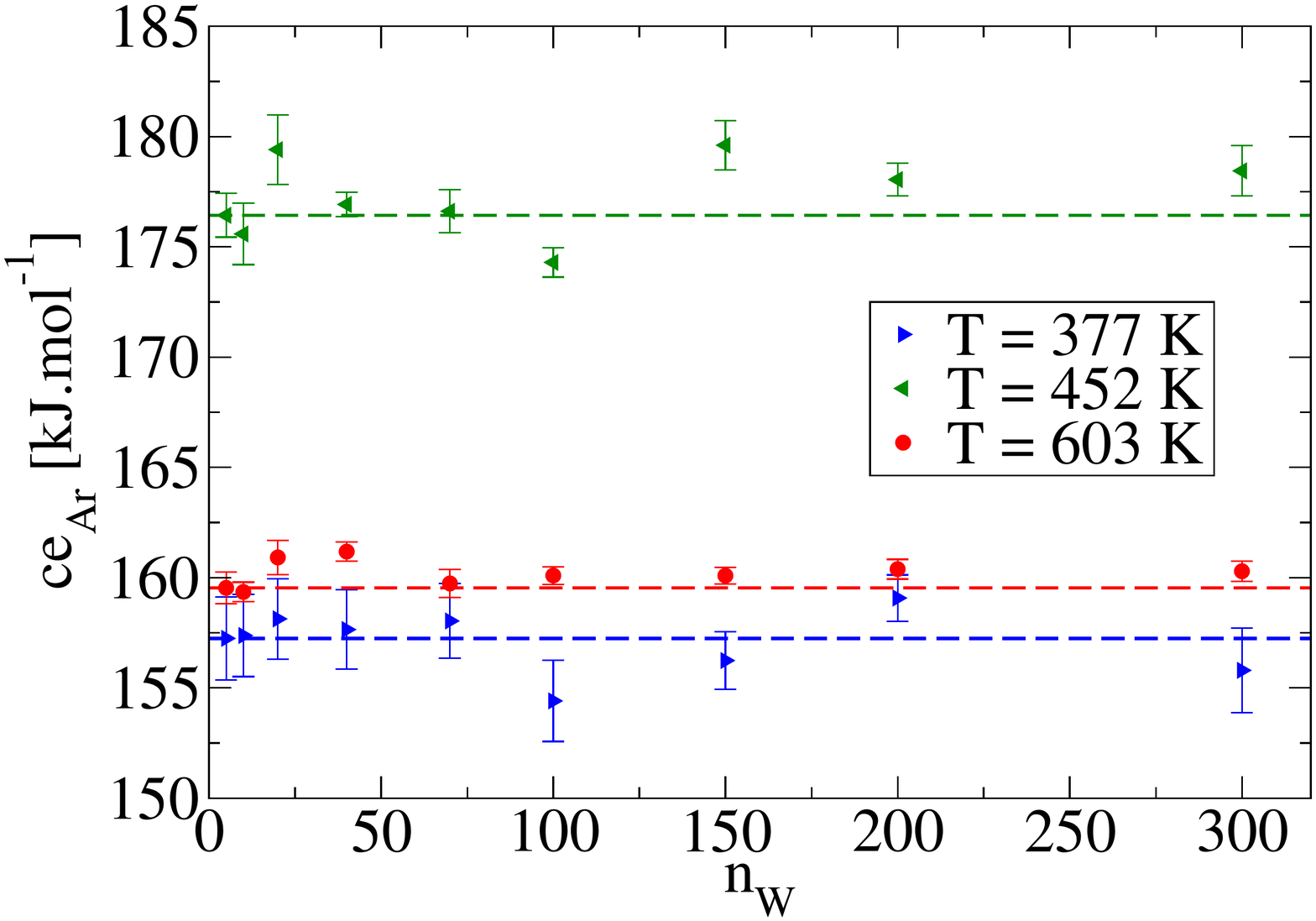}}
  \subfigure[]{\includegraphics[scale=0.25]{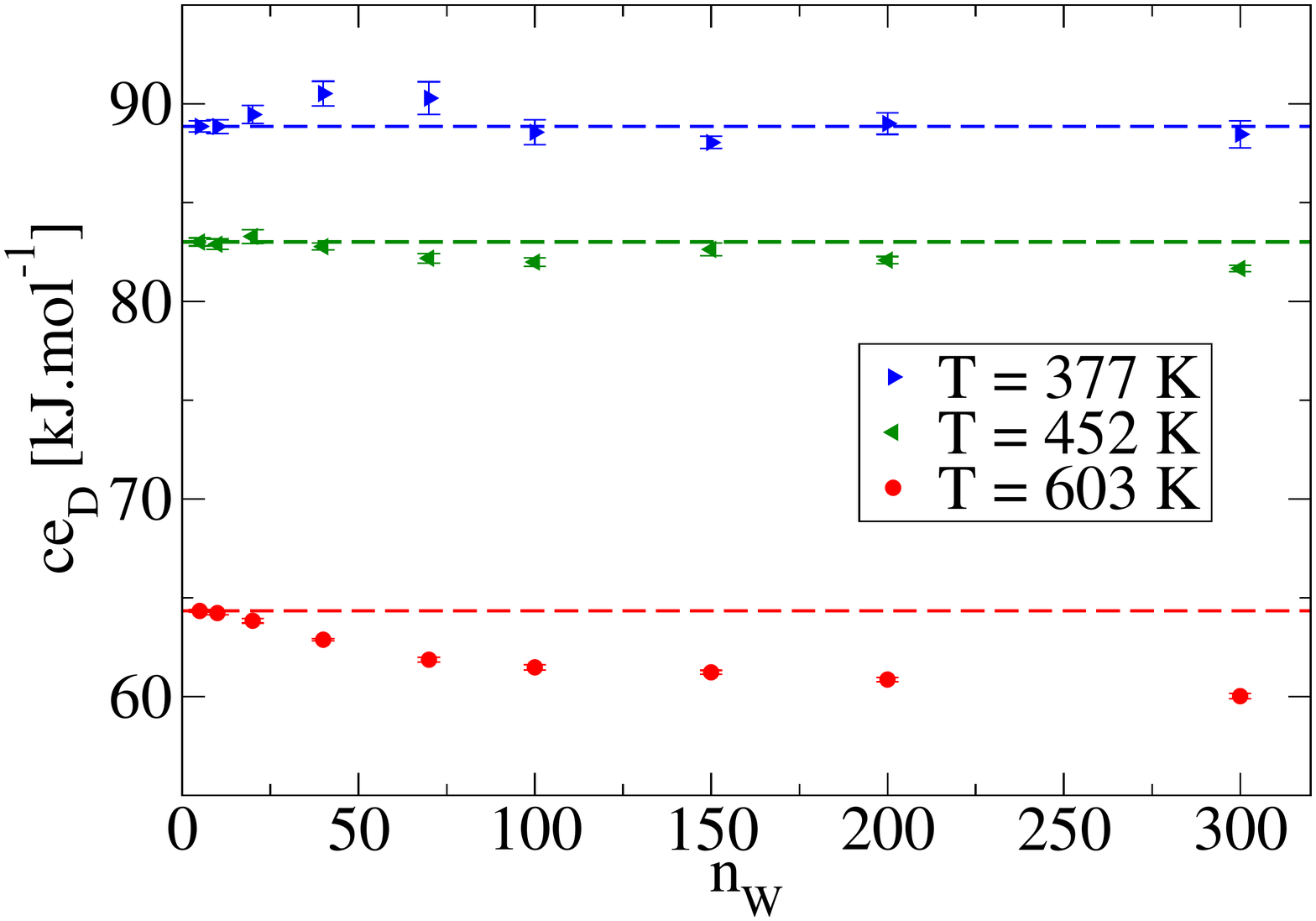}}
  \caption{\label{fig:potEnergy} (a): Variation of the cohesive energy density $ced$
with the number of water molecules $n_W$
for different temperatures.
(b): Variation of each contribution to the cohesive energy density as described
in Eq.~\ref{eq:cedContrib} with the number of water molecules for $T = 377$ K.
The dashed horizontal lines correspond to the value at $n_W = 5$.
(c) and (d): Variation of the cohesive energy between aromatic molecules $ce_{\text{Ar}}$
and docosane molecules $ce_{\text{D}}$, respectively, with the number of water molecules $n_W$
for different temperatures.
}
  \end{center}
\end{figure*}

Following the usual definition of the cohesive energy density~\cite{dack}
as the internal energy of vaporization 
of the liquid over the volume of liquid, we defined
the cohesive energy density of the water-bitumen mixture in the following way:
\begin{equation}
ced = -\frac{U^{\text{tot}} - U^{\text{intra}}}{V},
\end{equation}
where
$U^{\text{tot}}$ is the total potential energy in the system,
$U^{\text{intra}}$ the intramolecular potential energy between UAUs in the same molecule,
and $V$ the volume of the system.
The term $U^{\text{intra}}$ includes intramolecular bonding and non-bonding
interactions; adding the kinetic energy would lead to the internal
energy of the set of molecules as an ideal gas.
Figure~\ref{fig:potEnergy} (a) shows the variation of the cohesive energy density $ced$ with
the number of water molecules in the system for three different temperatures.
The values of the cohesive energy in the absence of water found
in the simulations are in agreement with experimental results on the
Hildebrand solubility parameter $\delta = \sqrt{ced}$ of bitumen.
Ref.~\cite{redelius} reports Hildebrand solubility parameters of bitumen
between $15.3$ (MJ/m$^3$)$^{1/2}$ and $23$ (MJ/m$^3$)$^{1/2}$, which
corresponds to a cohesive energy between $2.3$ and $5.3 \times 10^8$ Pa.
The simulation results lie exactly in this range.
Likewise, the decrease of the cohesive energy density with temperature
is very common.~\cite{barton}
Figure~\ref{fig:potEnergy} (a) shows clearly that, in the simulations, the cohesive energy density
 decreases with the water content.
This is so at all temperatures, although the trend is more visible
as the temperature increases.

This result can be investigated further. In particular, the 
contributions to the cohesive energy of the different molecule types
in each system
can be quantified for different water contents. 
The cohesive energy for a molecule of type $X$ is defined as
\begin{equation}
ce_{X} = -\frac{\Bigl(U_{X}^{\text{tot}} - U_{X}^{\text{intra}}\Bigr) \mathcal{N}_A}{N_{X} },
\end{equation}
where $N_{X}$ is the total number of molecules of type $X$ in the system,
$\mathcal{N}_A$ the Avogadro constant,
$U_{X}^{\text{tot}}$ the total potential energy between the molecules of type $X$ in the system,
and $U_{X}^{\text{intra}}$ the intramolecular potential energy of the molecules
of type $X$ in the system. Thus, the difference between these two energies corresponds to the intermolecular
energy between the molecules of type $X$ only.
The molecule type can be $\text{Ar}$ for aromatic, $\text{D}$ for docosane
(the saturates in the Cooee model), or $\text{W}$ for water. 
In the total cohesive energy density $ced$, the cross terms, corresponding to the intermolecular
energy between different molecule types, also matter.
They are defined here as:
\begin{equation}
ce_{X-Y} = -\frac{U_{X-Y}^{\text{inter}} \mathcal{N}_A }{(N_{X}+N_{Y})},
\end{equation}
where $U_{X-Y}^{\text{inter}}$ is the intermolecular energy between molecules of type $X$ and $Y$.
With these definitions, the total cohesive energy density $ced$ can be expressed as:
\begin{equation}
\label{eq:cedContrib}
ced = \frac{1}{V}\Bigl( \sum_X \frac{N_{X}}{\mathcal{N}_A} ce_{X} + \sum_{X \neq Y} \frac{N_{X} + N_{Y}}{\mathcal{N}_A} ce_{X-Y} \Bigr),
\end{equation}
where the sum $\sum_{X \neq Y}$ is done over all distinct pairs $(X,Y)$
such that $X \neq Y$.
Figure~\ref{fig:potEnergy} (b) shows the variation of the different contributions $ce_X$ and $ce_{X-Y}$
with the water content at $T = 377$ K.
The contributions involving water molecules, namely 
$ce_{\text{D}-\text{W}}$ and $ce_{\text{Ar}-\text{W}}$,
are growing with an increasing water content, while
$ce_{\text{W}}$ increases slightly. It means that 
the energetic contribution $N_{\text{W}} ce_{\text{W}}$ grows faster than linearly with an increasing water content.
This is not so surprising as the addition of one water molecule when only a few water molecules are present
causes the whole hydrogen bonded network to reorganise.
The other contributions to the cohesive energy stay constant. Consequently, the overall decrease of the cohesive energy
density $ced$ is not due to a decrease of the cohesive energy inside bitumen, but rather
to an increase in volume. In other words, when water molecules are added, the volume of the system increases,
some contributions to the cohesive energy increase,
but if bitumen molecules were added instead to match the same volume, the cohesive energy would increase more.
In total, the cohesive energy density decreases.
When the temperature increases, the picture is slightly different.
Figures~\ref{fig:potEnergy} (c) and (d) show the variation of the two cohesive energies $ce_{\text{Ar}}$
and $ce_{\text{D}}$, respectively, with the number of water molecules in the system, for
three different temperatures.
The cohesive energy $ce_{\text{Ar}}$ associated to aromatic molecules stays constant with the water
content at all temperatures. However, the cohesive energy $ce_{\text{D}}$ associated to docosane molecules
decreases with the water content at high temperatures. Thus, at high temperatures, the overall
decrease of the cohesive energy density $ced$ is due both to an increase in volume and to a decrease
of the cohesive energy in the docosane part of bitumen.
The distinction between docosane and aromatic molecules at high temperatures regarding their
intrinsic cohesive energy when water is added can be surprising.
Indeed, the aromatic molecules contain the slightly polar asphaltene
and resin molecules, which are usually believed to interact more with water~\cite{mullins2011, gray}
than with the apolar saturates.
If it was so, the water molecules should be located mainly close to the aromatic molecules,
replacing interactions among aromatic molecules by interactions between
water and aromatic molecules and thus decreasing the intrinsic cohesive energy of the aromatic molecules. 
This is not observed. Instead, it is the intrinsic cohesive energy of the docosane molecules
which is lowered, suggesting that water molecules are mainly close to docosane molecules and not to
aromatic molecules.


\begin{figure}
  \begin{center}
  \includegraphics[scale=0.3]{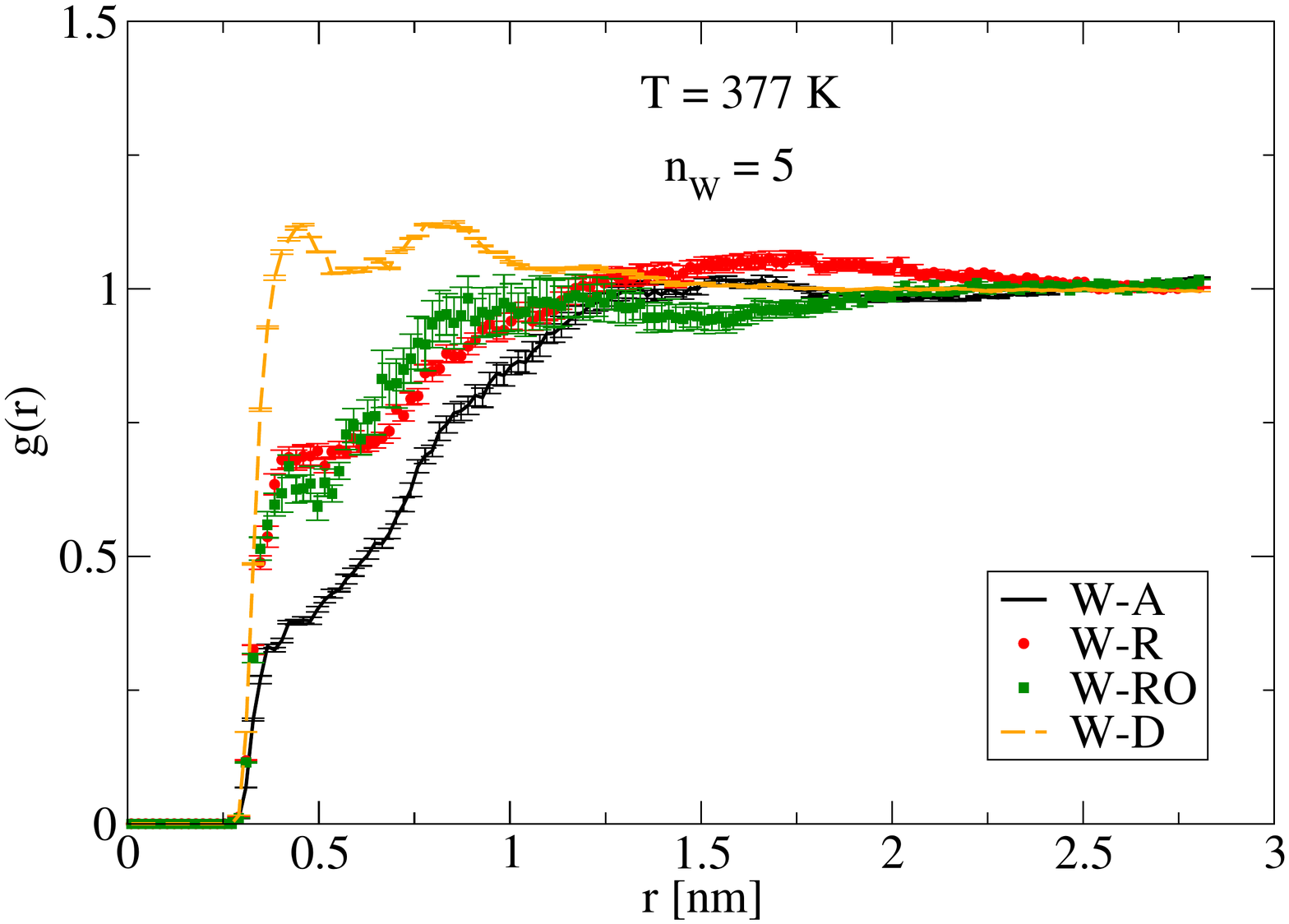}
  \caption{\label{fig:rdf} Radial distribution function between oxygen atoms in
 water molecules and other atoms
in other types of molecules for $n_W = 5$ at temperature $T = 377$ K.
Error bars correspond to standard deviation on the data divided by the
square root of the number of independent initial configurations considered.
}
  \end{center}
\end{figure}

The idea that water molecules are closer to the saturates than to the
polar aromatics can be directly checked
using radial distribution functions. The radial distribution functions
between the oxygen atom of the
water molecules and the atoms in other
molecules types
are displayed for the system with $n_W = 5$ water molecules and at
temperature $T = 377$ K in Fig.~\ref{fig:rdf}. This figure shows that
the atoms closest to oxygen atoms, except hydrogen atoms from other water molecules,
are from docosane molecules.
This result stays true for other temperatures and other water contents in
Cooee bitumen.
We believe that water molecules are closer to docosane molecules than to aromatic molecules
because aromatic molecules are part of the nanoaggregates.
Nanoaggregates are composed of aligned flat aromatic molecules in the simulations. They are also a
supramolecular structure identified experimentally in bitumen.~\cite{mullins2012}
The nanoaggregates are held together in the Cooee model by Lennard-Jones interactions
between many aligned united atoms. This can be seen as a model of the $\pi$-stacking
interaction existing between aromatic molecules in real bitumen.
The electrostatic interaction between
the positively charged hydrogen
of a water molecule and the slightly negatively charged
sulfur atom of an asphaltene or resin molecule is negligible
compared to the interaction holding the nanoaggregates together.
Thus, water molecules are unlikely to penetrate the nanoaggregates
and have no other choice than to stay close to the docosane molecules.
We would like to stress here that this result could be altered if the
polarity of the asphaltene and resin molecules is increased enough,
or if they have the possibility to form hydrogen bonds. However, the
molecular structures used in this work for asphaltene and resin molecules
are believed to be representative.

\begin{figure}
  \begin{center}
  \includegraphics[scale=0.3]{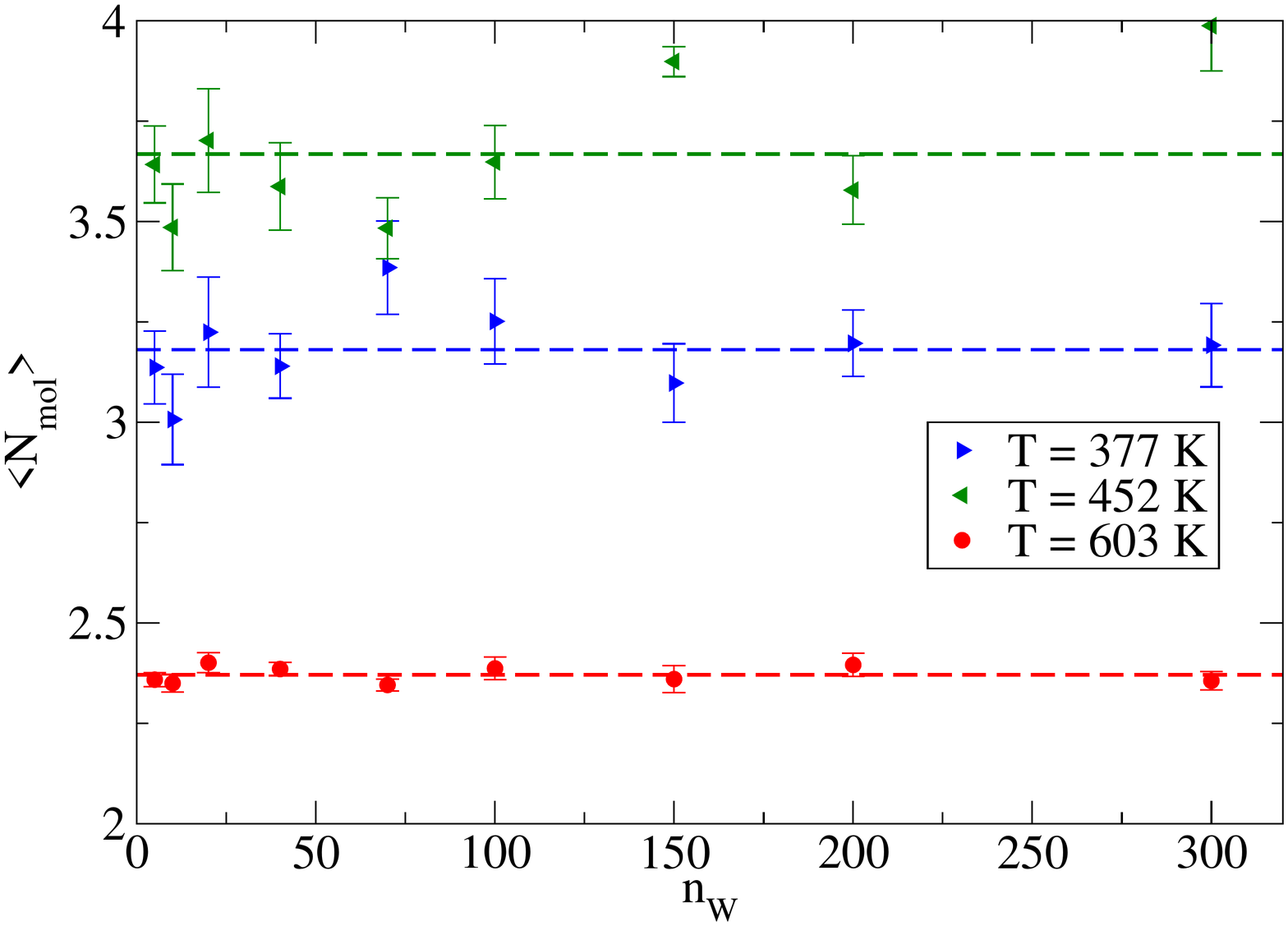}
  \caption{\label{fig:clusterSize} Variation of the average number of aromatic molecules $\langle N_{\text{mol}}\rangle$
 with the number of water molecules $n_W$
for different temperatures. The dashed horizontal lines correspond to the average value
over all values of $n_W$ for the same temperatures.}
  \end{center}
\end{figure}

The variation of the intrinsic energy of the aromatic molecules $ce_{\text{Ar}}$
with temperature and water content deserves some further discussion.
As already mentioned, the aromatic molecules in the Cooee bitumen are known to align in nanoaggregates.
The definition
of the nanoaggregates in the case of Cooee bitumen is detailed in
Refs.~\cite{aging, branched} 
It sums up to the following rule:
two aromatic molecules
are nearest neighbors in the same nanoaggregate if they are
well aligned and close enough.
More specifically, this rule is based on three thresholds.
The first threshold quantifies how much the molecules should be aligned
to be declared in the same nanoaggregate.
This first threshold imposes boundaries to the angle $\theta$ between the normal vectors to the aromatic planes
of two different molecules. We choose $0\degree \leq  \theta \leq 34\degree$
and $149\degree \leq \theta \leq 180\degree$.~\cite{aging}
The second threshold is related to how close the molecules should be to be declared
in the same nanoaggregate.
This second threshold imposes a maximum value to the distance $d_1$ between the aromatic planes
of two different molecules. We impose $d_1 \leq 6$ \AA$ $.~\cite{aging}
Two molecules far away but in the same plane can have a very low distance $d_1$. This is
why a third threshold is needed.
The third threshold imposes a maximum value to the distance $d_2$ between the center of mass
of the first molecule and the projection
of the center of mass of the second molecule on the plane of the first molecule.
We fix $d_2 \leq 0.7d_A$ \AA$ $,~\cite{aging} where $d_A=13.1$ \AA\  is the typical length
of an asphaltene molecule
in the Cooee model.
Some nanoaggregates are branched, because the asphaltene molecule
chosen in the Cooee model has a flat head and a flat body oriented in different directions, and
both parts can align with other aromatic molecules. 
The average number of aromatic molecules inside a nanoaggregate is used to quantify
the size of the nanoaggregates.
The variation of the nanoaggregate size with the water content is displayed in Fig.~\ref{fig:clusterSize}
for three different temperatures.
This figure shows that within error bars and for all temperatures, the nanoaggregate
size does not depend on the water content. This is in agreement
with the intrinsic cohesive energy of the aromatic molecules being independent of the water content.
Surprisingly, the variation of the nanoaggregate size with temperature is non-monotonic.
This fact has been reported and explained.~\cite{branched} It is due to two competing effects
when temperature increases: the first effect is the increase of thermal noise which tends to detach
molecules from the nanoaggregates and decrease the nanoaggregate size; the second effect is the
relative increase of the number of asphaltene molecules inside the nanoaggregates compared to other molecule
types, which tends to increase the degree of branching of the aggregates and also their size.
The non-monotonic behavior of the cohesive energy $ce_{\text{Ar}}$ between aromatic
molecules is due to the same cause.

\begin{figure}
  \begin{center}
  \includegraphics[scale=0.3]{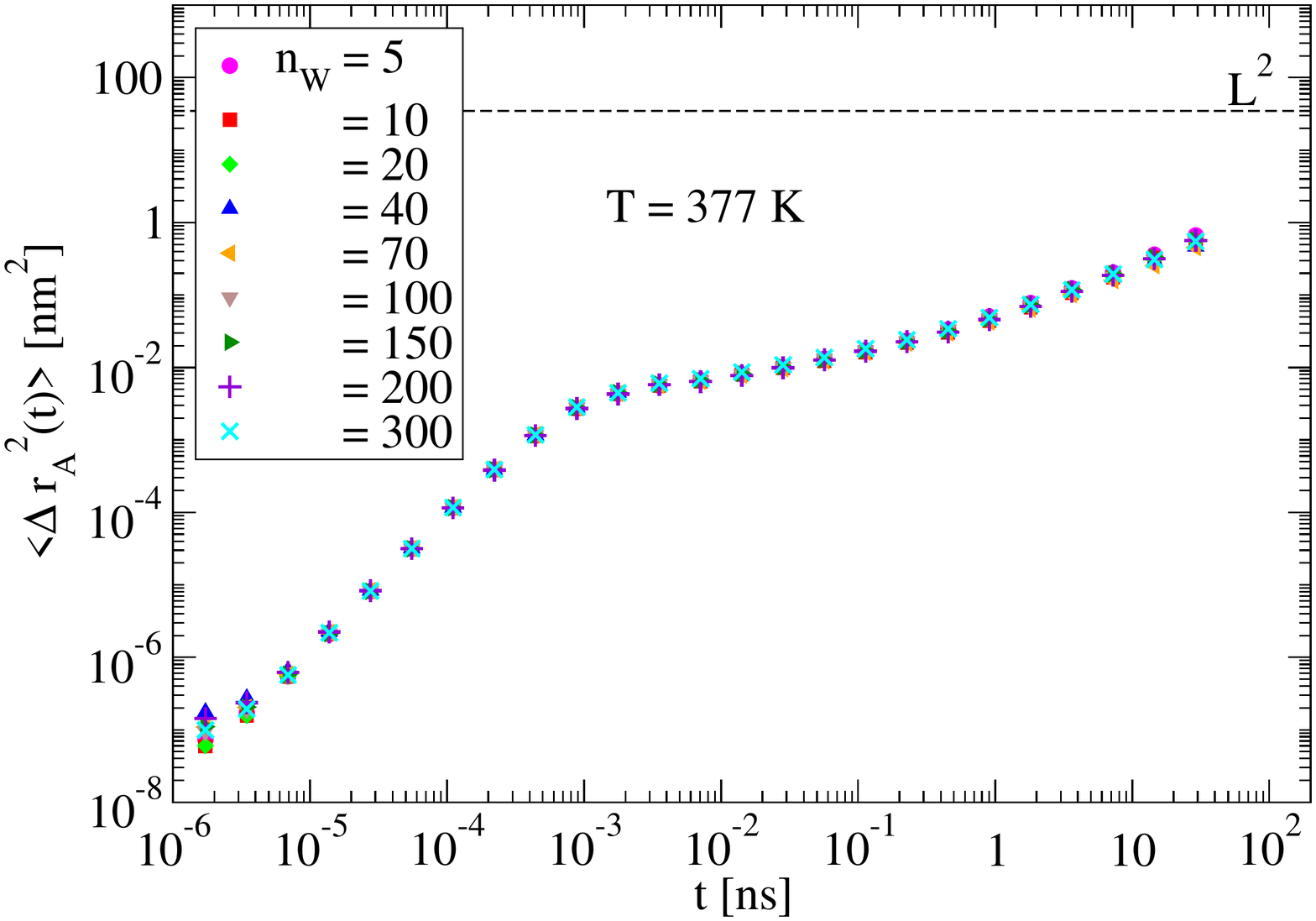}
  \caption{\label{fig:msdAsph} Time evolution of the mean-squared displacement
of the center of mass of the asphaltene molecules for different numbers of water molecules $n_W$
at temperature $T = 377$ K.
The dashed black line indicates the value of the box length squared ($L^2$),
in the case of of $n_W = 5$.
}
  \end{center}
\end{figure}

Finally, the variation of the bitumen dynamics with the water content can be checked.
Figure~\ref{fig:msdAsph} displays the time evolution of the mean squared displacement
of the centers of mass of the asphaltene molecules for different
water contents and at temperature $T = 377$ K.
The shape of the curves is characteristic of that of a viscous liquid.
At short time scales a ballistic regime is visible. It is followed
by a plateau at intermediate times and finally a diffusive regime is recovered
All curves collapse, which shows that the asphaltene dynamics do not depend on the water content.
The same result is found for the dynamics of the saturates molecules (not
shown).

\subsection{Water droplet}

\begin{figure}
  \begin{center}
    \includegraphics[scale=0.35]{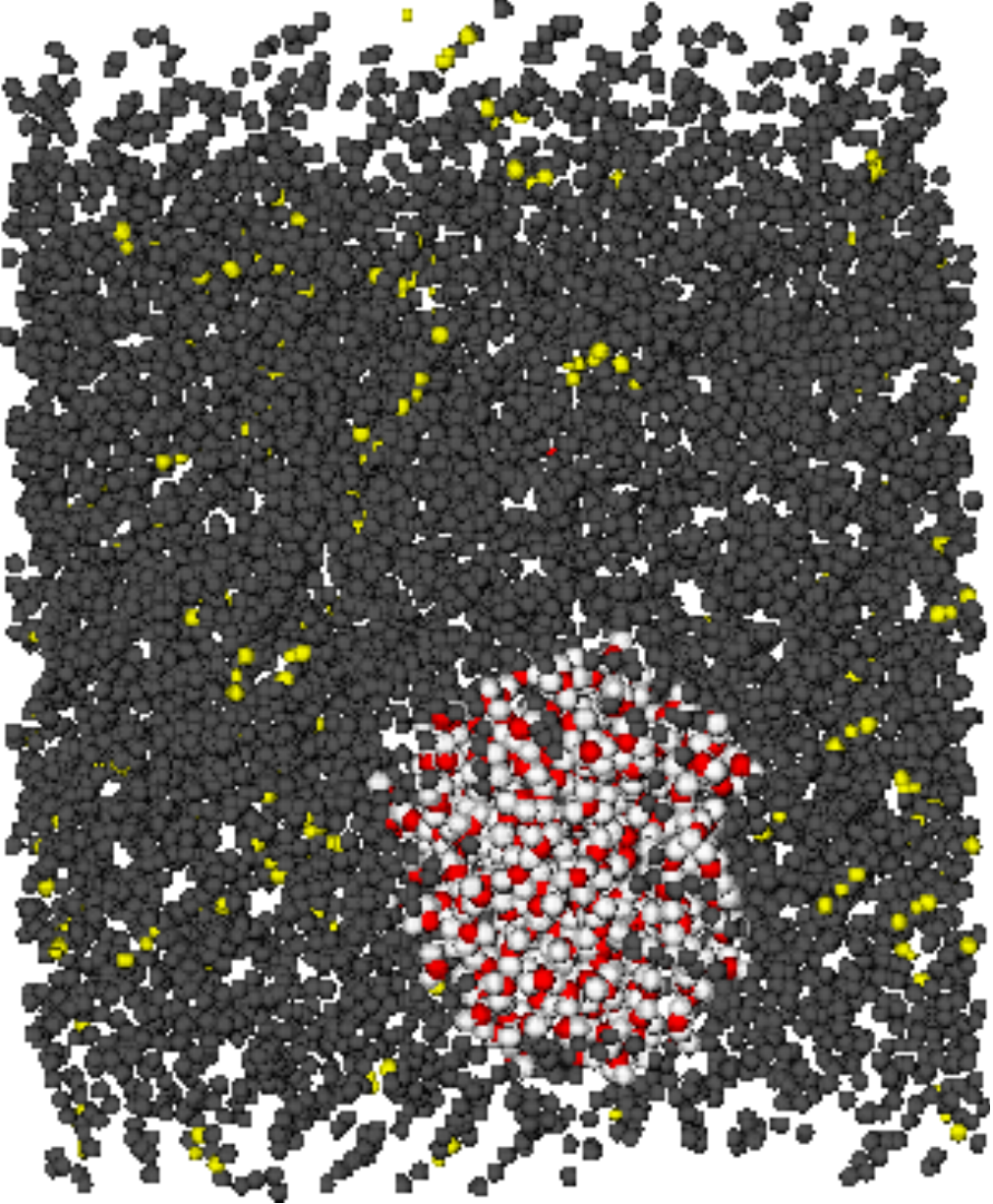}
    \caption{\label{fig:dropImage} (Color online)
      Snapshot
      of the $n_w=300$ system at $T=452$ K. All carbon-based UAU
      are grey, sulfur atoms are yellow, hydrogen white and oxygen red.
      }
  \end{center}
\end{figure}

The water molecules present in bitumen generally form a droplet,
as can be seen in Fig.~\ref{fig:dropImage} for 
one configuration of the system with $n_W = 300$
water molecules and at temperature $T = 452$ K. 
The water distribution among droplets and the droplet sizes were quantified by
Delaunay tetrahedra for which at least 2 of 4 vertex atoms belonged to a water
molecule.

\par
\begin{figure}
  \begin{center}
  \scalebox{0.5}{
   \includegraphics{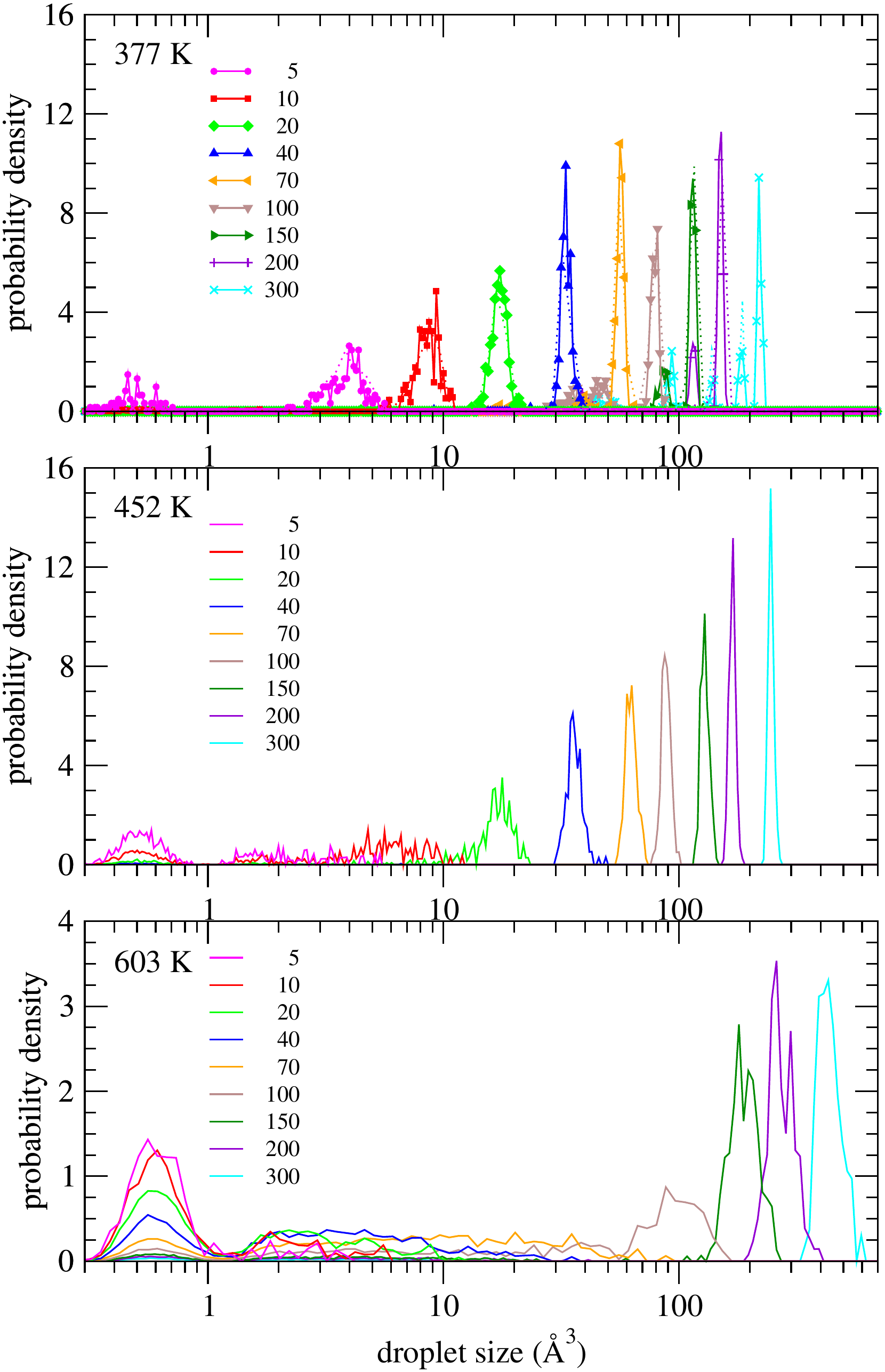}
   }
    \caption{\label{fig:drop_Vdistrib} (Color online)
      Volume-weighted distribution of droplet volumes at $T=377$, 452, and 603~K.
      Separate distributions are computed for each number of water molecules.
      Dotted lines indicate results at 377~K using only the final
          1/3 of each trajectory.
      }
  \end{center}
\end{figure}

The droplet volume was approximated as the volume of adjoining tetrahedra that
form a cluster.  The distribution of droplet volumes under all conditions are
shown in Fig.~\ref{fig:drop_Vdistrib}.  The y-axis indicates the probability
that a volume $V$ of water can be found within droplets of volume between $V$
and $V + dV/V$;
it is normalized so $\int P(V) d\ln V = 1$.  Results at 377~K
show that a range of droplet sizes can occur.  While most volume is usually
found in a single large droplet, significant numbers of cases were found with
multiple droplets with volumes between 90 and 200~\AA$^3$ for the $n_w=300$
system.  Similar results were found for $n_w = 200$ to 70 water molecules.
Only a single large droplet was typical for systems with fewer than 70 water
molecules.

\par 
Results at 452~K differ in that only a single droplet size was prevalent at
each composition.  This potentially reflects better equilibration of the
bitumen--water phase behavior due to faster water diffusion over nanosecond
time scales (see below).

\par 
Results at 603~K show a qualitative difference.  A broad distribution of drop
sizes over a 1 to 70~\AA$^3$ range is present at all water compositions.  For
$n_w\le 40$, most volume is found within states of $V < 1$~\AA$^3$.  These
correspond to individual water molecules.  
Coalescence into larger droplets occurred for $n_w = 100$ and larger, though
more water molecules remained outside the largest droplet than in the 377 and
452~K cases.

\par
\begin{figure}
  \begin{center}
  \scalebox{0.8}{
  \includegraphics{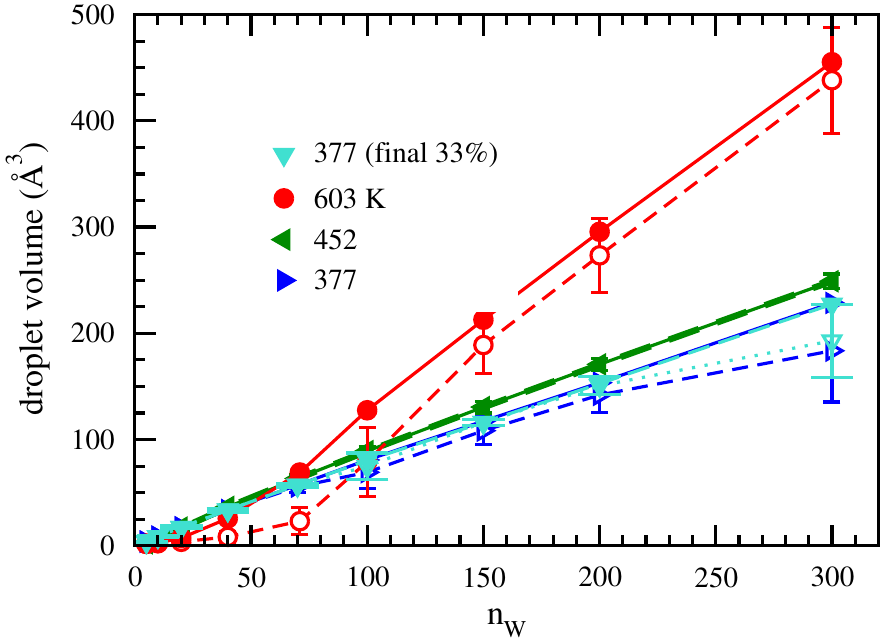}
  }
    \caption{\label{fig:drop_Vavg} (Color online)
      Average volume of the largest droplet (open symbols, dashed lines) and
      of all large droplets (closed symbols, solid lines) at temperatures 377,
      452, and 603~K and at 377~K using the final 1/3 of the trajectory.
      }
  \end{center}
\end{figure}

 
Two measures of the droplet volume are shown in Fig.~\ref{fig:drop_Vavg}.
Solid lines indicate the total volume of Delaunay tetrahedra that constitute
large water droplets ($V\ge 1$~\AA$^3$).  Dashed lines indicate the volume of only the largest
droplet. Error bars indicate standard deviations on the size of the largest
droplet present, averaged over all configurations considered.

The total droplet volume increases approximately linearly with the number of
water molecules in the system.  Deviations from linearity are most notable at
the highest temperature for the smaller numbers of water molecules,
i.e.\ the cases that are dominated by an absence of coalesced droplets.
Linearity indicates a homogeneous water environment as the number of water
molecules increases.

The average volumes of only the largest droplets indicate differences among
the three temperatures that are consistent with the volume distributions.
Many drops are present in the 603~K
case, leading to the largest drop containing only a small to moderate fraction
of the total drop volume until a large number of water molecules are present
(mass fraction of ca.~1.6\%).  Some cases of multiple droplets at
377~K lead to similar but much smaller effects at that temperature at larger
numbers of water molecules. The larger error bars at high $n_W$ for both
  temperatures are a consequence of the volume differences for the largest
  droplet  when 2 or more droplets are present, such as 220~\AA$^3$
  vs.\ 180~\AA$^3$ in the case of 1 or 2 droplets at 377~K with $n_W=300$ 
  (see Fig.~\ref{fig:drop_Vdistrib}).
Results at 452~K show a more predominant
occurrence of a single large droplet. Cases of multiple large droplets
were sufficiently rare at this temperature to lead to negligible
differences between the average
volume of the largest droplet and of all large droplets.

%


\par
\begin{figure}
  \begin{center}
  \scalebox{0.8}{
      \includegraphics{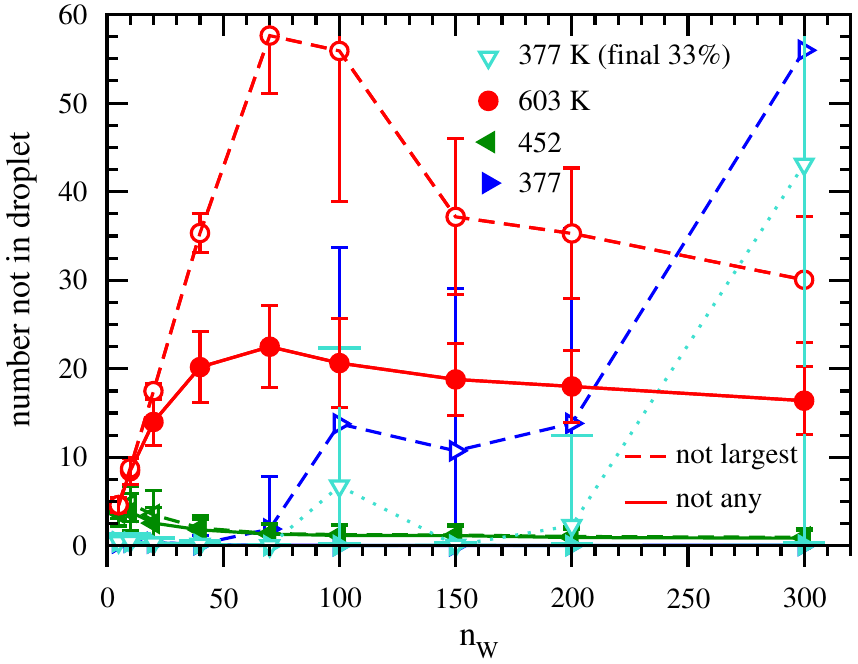}
  }
    \caption{\label{fig:drop_fracW} (Color online) Number of water molecules
      not in the largest droplet (dashed line, open symbols) or not in any
      droplet (solid line, filled symbols) at temperatures 377, 452, and
      603~K and at 377~K using the final 1/3 of the trajectory.
      }
  \end{center}
\end{figure}

The division of water molecules between large droplets and dispersed molecules
was also determined from the geometric analysis.
The number of water molecules that are not
in droplets of $V > 1$~\AA$^3$ are shown in Fig.~\ref{fig:drop_fracW} by
solid lines. It corresponds to the number of free water molecules.
Error bars indicate standard deviations across all configurations.
Essentially all water molecules
are in droplets at 377~K, and the number of free water molecules increases
with temperature. At 377 and 452~K, the number of free water
molecules decreases as the total number of water molecules increases 
beyond $n_W=10$ in the
simulations. 
At 603~K, most water molecules are free for concentrations up to
$n_W=40$, then the number reaches a peak and decreases as the water
concentration increases further.  These results indicate that additional water
molecules add to the largest droplet.  This is consistent with a shift in the
droplet size distribution to larger volumes, as seen in Fig.~\ref{fig:drop_Vdistrib}.

Dashed lines indicate the number of water molecules that are not in the
largest droplet. Error bars indicate the standard deviation in the
  number of water molecules that {\em are} in the largest droplet; it applies
  to counting those not in the largest droplet because the number of molecules
  is constant in each simulation.
Results at 603~K differ qualitatively by showing a large
fraction of water molecules at all concentrations that are not in 
the largest droplet.  This constitutes more than half of the water molecules 
for systems with up to 100 water molecules. Results at 452~K show
  essentially all water molecules belonging to a single large droplet.

\par
Differences between 377 and 452~K are suggested by the results
at higher numbers of water molecules.  Multiple droplets exist at 377~K
  for $n_W=100$ and larger when averaged over the entire simulation trajectory,
  which is consistent with the multiple peaks shown in Fig.~\ref{fig:drop_Vdistrib}. 
  This leads to a significant number of water molecules outside of the
  largest droplet.  However, further analysis shows that when multiple
    droplets exist in cases of $n_W=40$ or more, they
 are present initially and coalesce into fewer droplets as
  the simulation proceeds. This suggests that a longer
  equilibration period would lead to one large droplet being predominant at
  377~K, as is found at 452~K.

\par 
To test this idea, the 377~K results were reanalyzed by using only the final
one third of the structures in each production run.  Results are shown by
dotted lines.  The volume of the largest droplet (Fig.~\ref{fig:drop_Vavg})
shifts toward the volume of all large droplets as a consequence of 
droplets coalescing.  Similarly, the number of water molecules not in the
largest droplet decreases (Fig.~\ref{fig:drop_fracW}).  Changes in
the size distributions (Fig.~\ref{fig:drop_Vdistrib}) are less apparent, in
part because of the wide volume scale.  A clear change is the loss of droplet
volumes between 30 to 50~\AA$^3$ at $n_W=100$.  Droplets of this size coalesce
with the main droplet and do not form again during the simulation.  These results emphasize the long
time scales required for structural relaxations at lower temperatures.

\subsection{Water dynamics}

\begin{figure*}
  \begin{center}
  \subfigure[]{\includegraphics[scale=0.25]{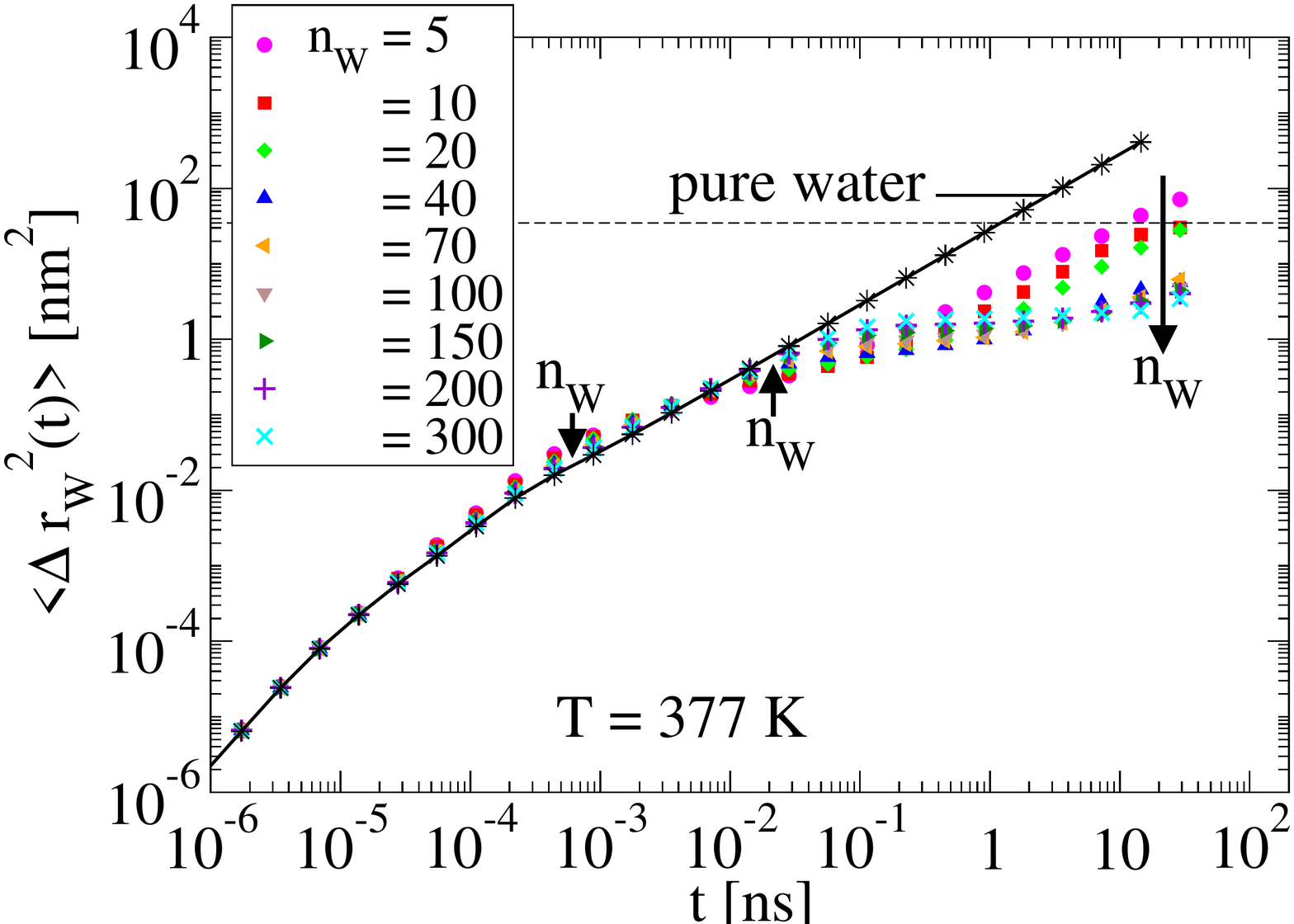}}
  \subfigure[]{\includegraphics[scale=0.25]{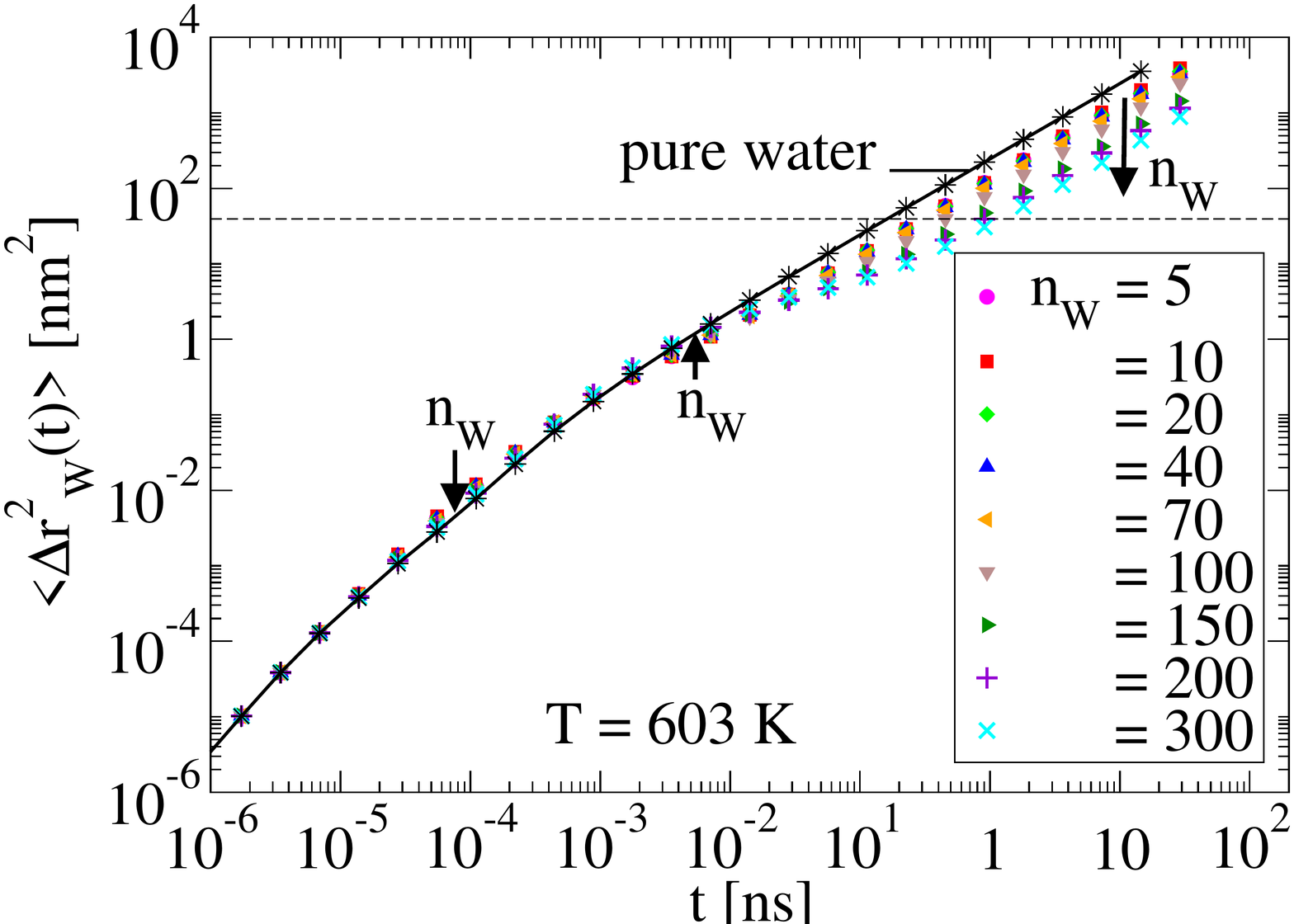}}\\
  \subfigure[]{\includegraphics[scale=0.25]{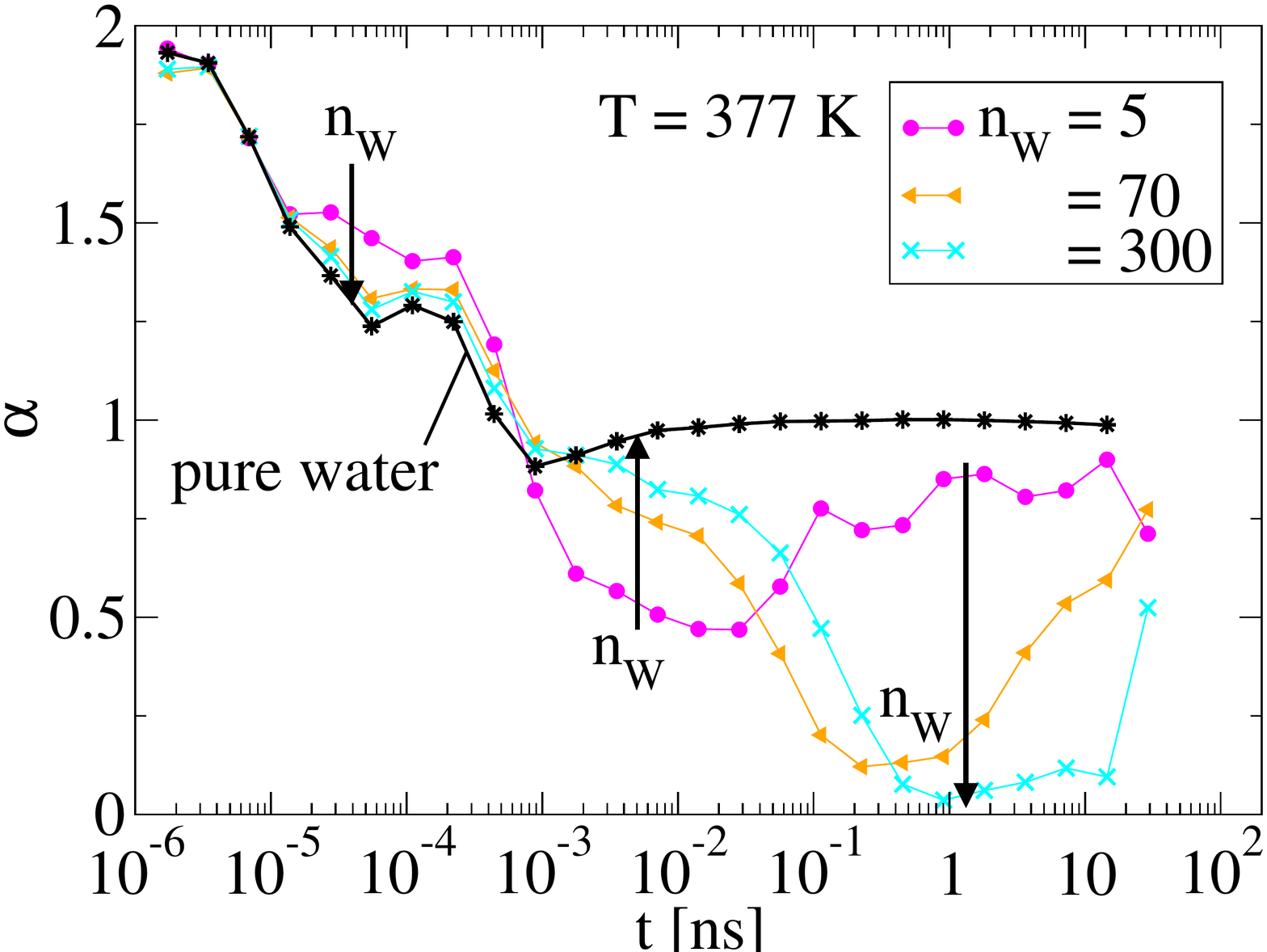}}
  \subfigure[]{\includegraphics[scale=0.25]{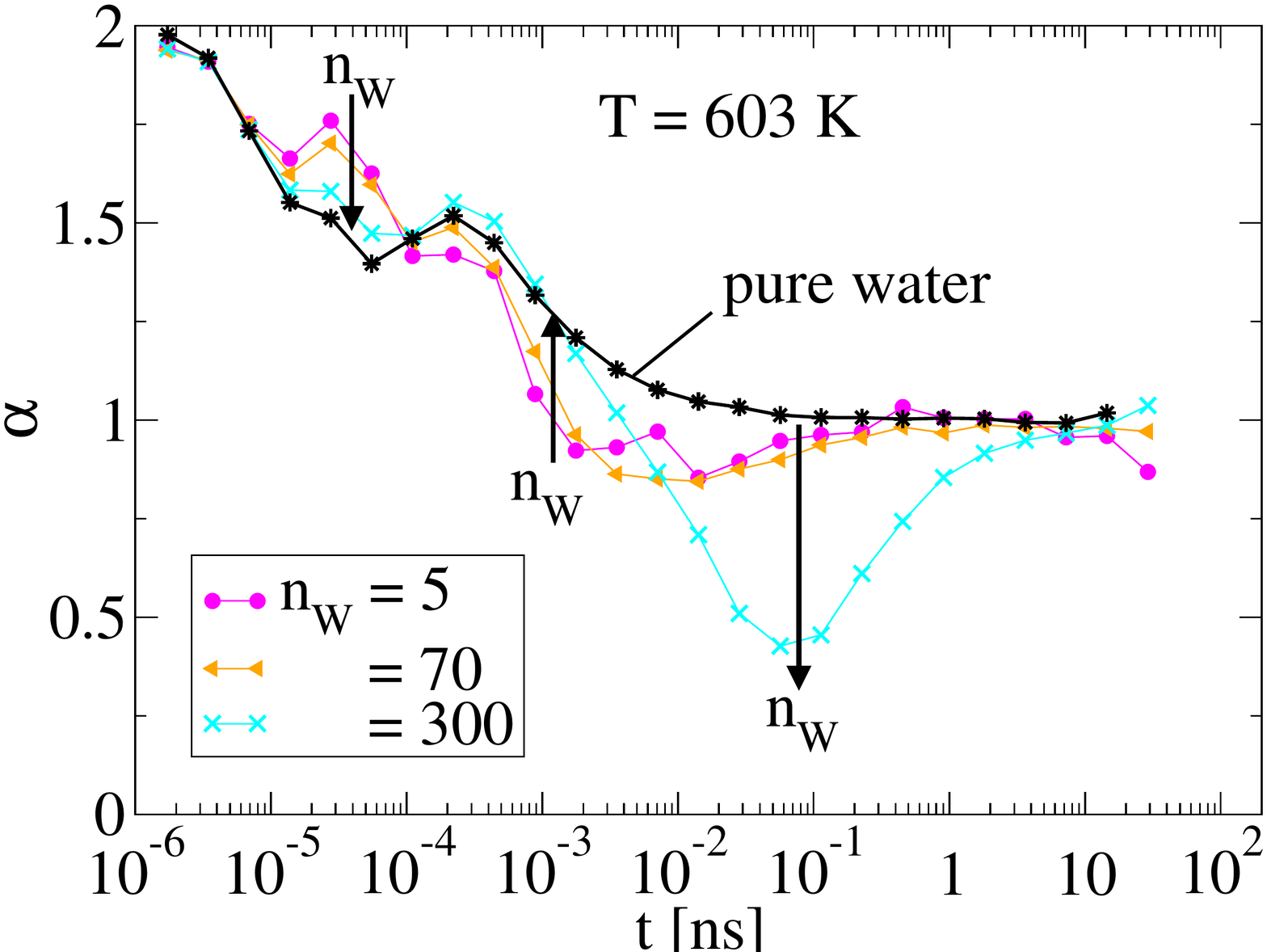}}
  \caption{\label{fig:msdW} Variation of the mean squared
displacement $\Delta r_W^2(t)$ of the center of mass of water molecules
with time for different number of water molecules and for temperatures $T = 377 K$  (a)
and $T = 603 K$ (b).
The black dashed line indicates the value of the box length squared,
in the case of of $n_W = 5$.
(c) and (d): Variation of the exponent of the mean squared displacement curves in (a)
and (b), respectively.
For the sake of readability only curves corresponding to $n_W = 5, 70, 300$ are displayed.
}
  \end{center}
\end{figure*}

\begin{figure}
  \begin{center}
  \includegraphics[scale=0.3]{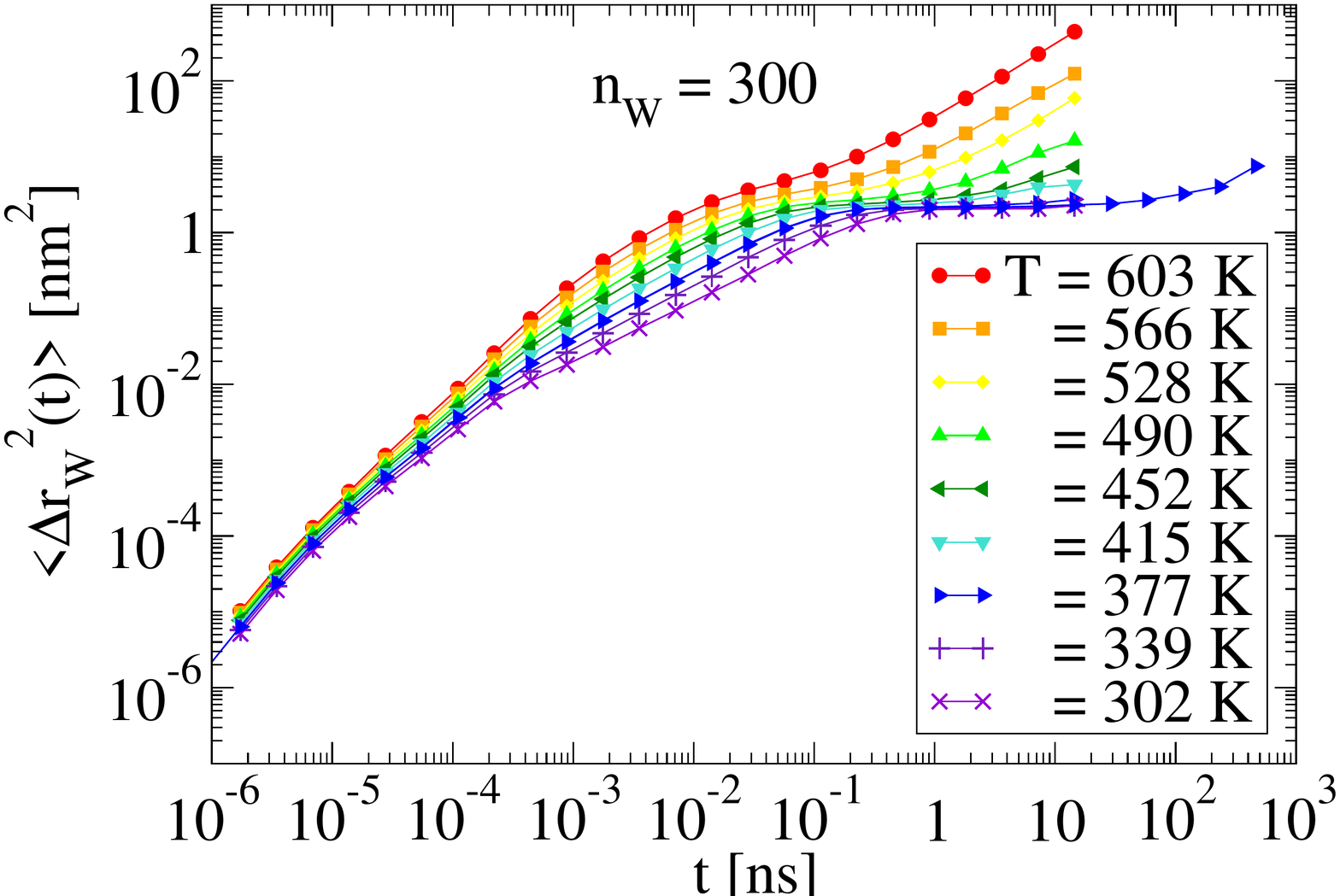}
  \caption{\label{fig:msdW300} Variation of the mean squared
displacement $\Delta r_W^2(t)$ of the center of mass of water molecules
with time for $n_W = 300$ at different temperatures.}
  \end{center}
\end{figure}

The water dynamics are affected by the fact that water molecules tend to form a droplet.
The water dynamics are first quantified using the mean squared
displacement $\Delta r_W^2$
of the centers of mass of the single water molecules. This mean squared
displacement is plotted versus time for different water contents
and for temperature $T=377$ K and $T = 603$ K, in Figs.~\ref{fig:msdW} (a) and (b),
respectively.
This figure also displays the mean squared displacement of
pure water at the same temperature and pressure.
The main result of this figure is that the water dynamics is slowed
down significantly at long times in bitumen compared to its counterpart in pure water.
This is true at all temperatures, although the slow down is
larger at low temperatures.
Surprisingly, the slow down is larger for large water contents
than for low water contents.
A diffusive regime is eventually reached at long times. The long-time diffusive regime is clearly visible
at high temperatures and small water contents 
in Figs.~\ref{fig:msdW} (a) and (b). It is also observable
at the low temperature $T=377$ K
and for the largest water content $n_W = 300$, for a long enough simulation.
This is shown in Fig.~\ref{fig:msdW300}, displaying in particular
the time evolution of the mean squared displacement of water molecules
in a single simulation of $0.86$ $\mu$s with $n_W = 300$ water molecules
at $T=377$ K.
The diffusion constant of the long-time diffusion regime decreases with an increasing
water content. This indicates that this regime corresponds to the diffusion of the
water droplet inside the bitumen matrix. 
As the number of molecules increases, the droplet becomes heavier and
its diffusion constant decreases.
Of course, in the limit of many water molecules with a droplet
of bitumen in it, the faster bulk water dynamics would be recovered, but this
is not the limit studied in this work.

Two other interesting observations can be made on the
time evolution of the mean squared displacement of water molecules
at shorter time scales.
The first observation is that the water dynamics in bitumen are slightly faster than
the pure water dynamics at short times, between $10^{-4}$ and $10^{-3}$ ns, as
can be seen in Figs.~\ref{fig:msdW} (a) and (b).
The smaller
the number of water molecules, the faster the dynamics at these time scales.
The second observation is that at intermediate times, around $10^{-2}$ ns,
the water dynamics in bitumen becomes slower than in pure water, but the larger
the number of molecules in bitumen, the closer to the bulk water dynamics.
To interpret these observations, it is necessary to describe the time evolution
of the mean squared displacement of water molecules more closely.
An informative quantity in this respect is
the local exponent $\alpha(t)$ of the mean squared displacement curve.
It is defined as
\begin{equation}
\alpha (t) = \frac{\log{\Delta r_W^2(t)}-\log{\Delta r_W^2(t-\Delta t)}}{\log{(t)}-\log{(t-\Delta t)}},
\end{equation}
where $\Delta t$ is increasing logarithmically in base $2$.
The time evolution of the local exponent $\alpha$ is displayed in Figs.~\ref{fig:msdW} (c) and (d)
for temperatures $T = 377$ K and $T = 603$ K, respectively. 
For the sake of clarity, the case of temperature $T = 377$ K is described in more details,
results being qualitatively similar at other temperatures.
In Fig.~\ref{fig:msdW} (c), the ballistic motion is clearly visible
at the beginning and corresponds to $\alpha = 2$.
At time scales around
$10^{-4}$ ns, the local exponent $\alpha$ is smaller than $2$ and 
decreases more as the number of water molecules increases.
In other words, when only a few water molecules
are present in bitumen, it is as if the ballistic regime was maintained for longer times.
This could be due to the fact that, as the number of water molecules decreases, water molecules 
interact with more bitumen molecules, with which the interaction is lower than with other
water molecules, maintaining a ballistic
motion for longer times.
It explains why the water dynamics around $10^{-4}$ ns is slightly faster when the
number of molecules in bitumen is lower, as seen in Fig.~\ref{fig:msdW} (a).
Around $10^{-3}$ ns, the regime becomes diffusive in bulk water,
which is characterized by $\alpha = 1$. It corresponds to diffusion of water molecules
in water. At this time scale, the regime becomes
subdiffusive for water molecules in bitumen. The smaller the number of water molecules
in bitumen, the earlier the start of the subdiffusive regime. For example,
for $n_W = 5$ water molecules, the diffusive regime seen in bulk water is
not reached and a subdiffusive regime settles in at $10^{-3}$ ns, whereas
for $n_W = 300$ the diffusive regime of bulk water is followed up to $5\times10^{-3}$ ns.
This effect is likely due to water molecules feeling the edge of the water droplet.
The bigger the droplet, the later this effect is observed. 
It explains why at a given time scale around $5\times10^{-2}$ ns in Fig.~\ref{fig:msdW} (a)
the dynamics of water molecules is faster for a larger number of water molecules: they are not yet
affected by the drop edge.
At longer time scales and as already discussed, the water dynamics is governed by the diffusion
of the water droplet inside bitumen.
For large water contents, the final diffusive regime sets in after a long plateau,
associated with a local exponent $\alpha$ close to zero,
and corresponding to the water droplet being nearly arrested in bitumen.

\begin{figure}
  \begin{center}
  \subfigure[]{\includegraphics[scale=0.25]{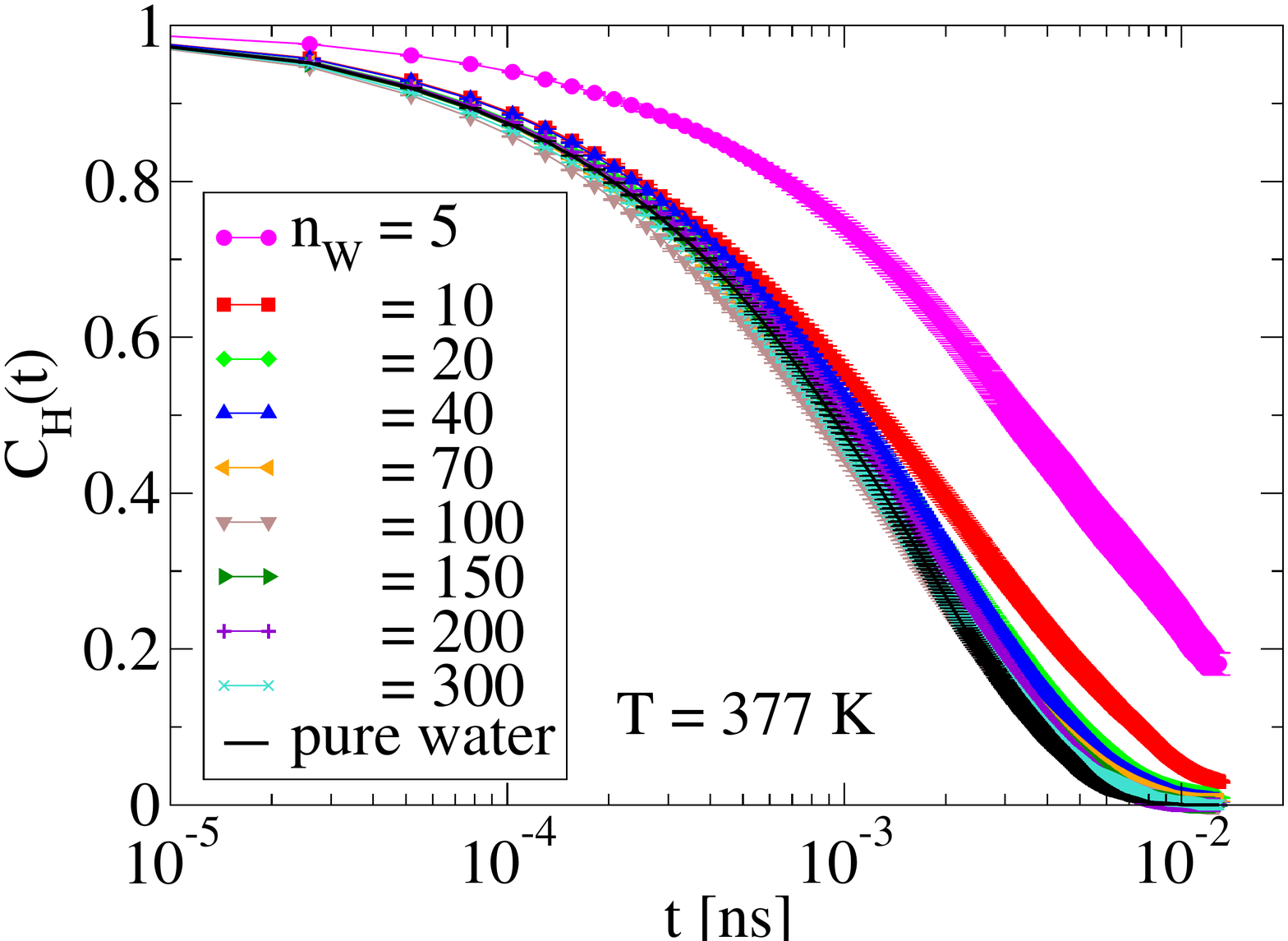}}
  \subfigure[]{\includegraphics[scale=0.25]{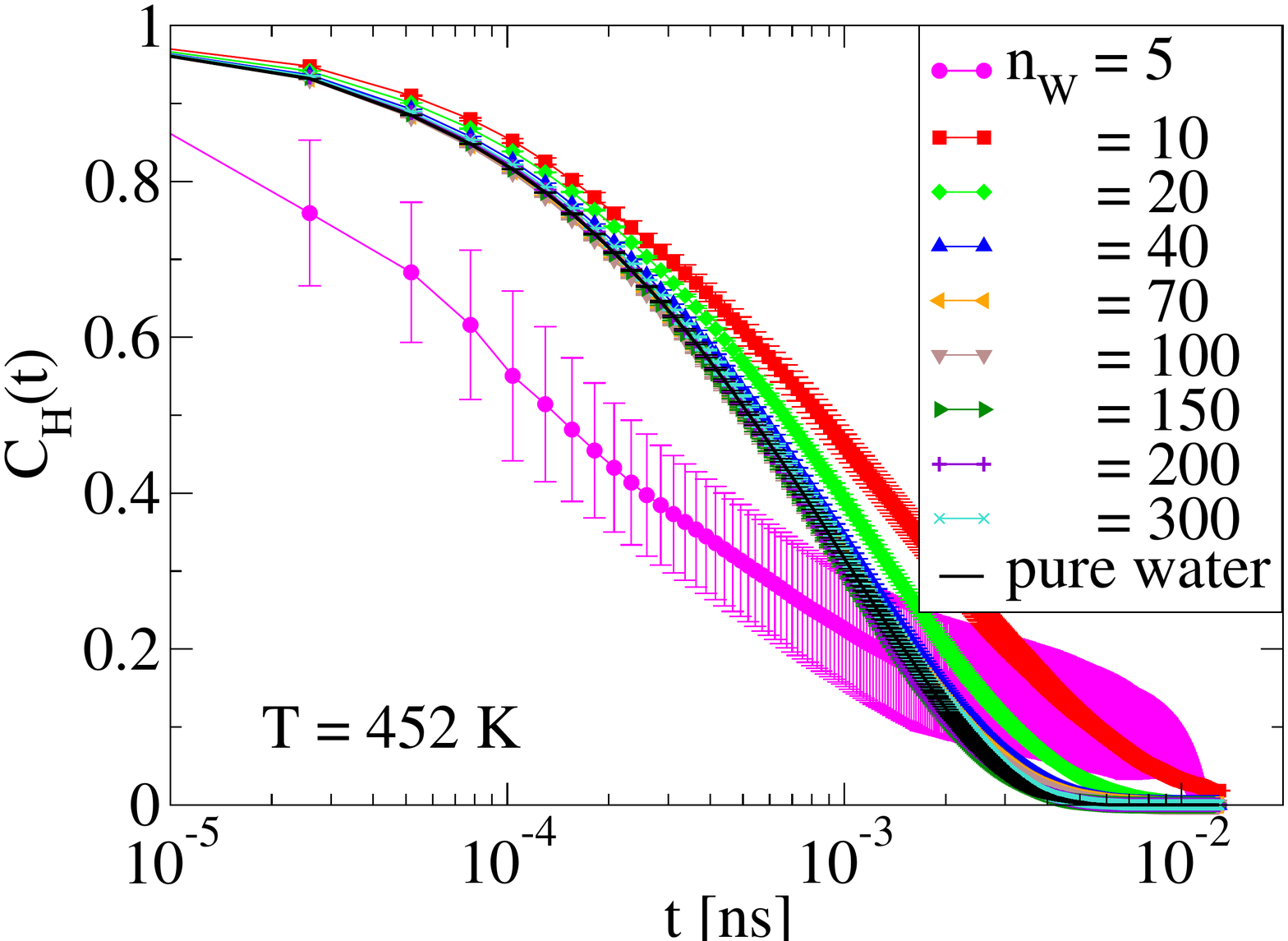}}\\
  \subfigure[]{\includegraphics[scale=0.25]{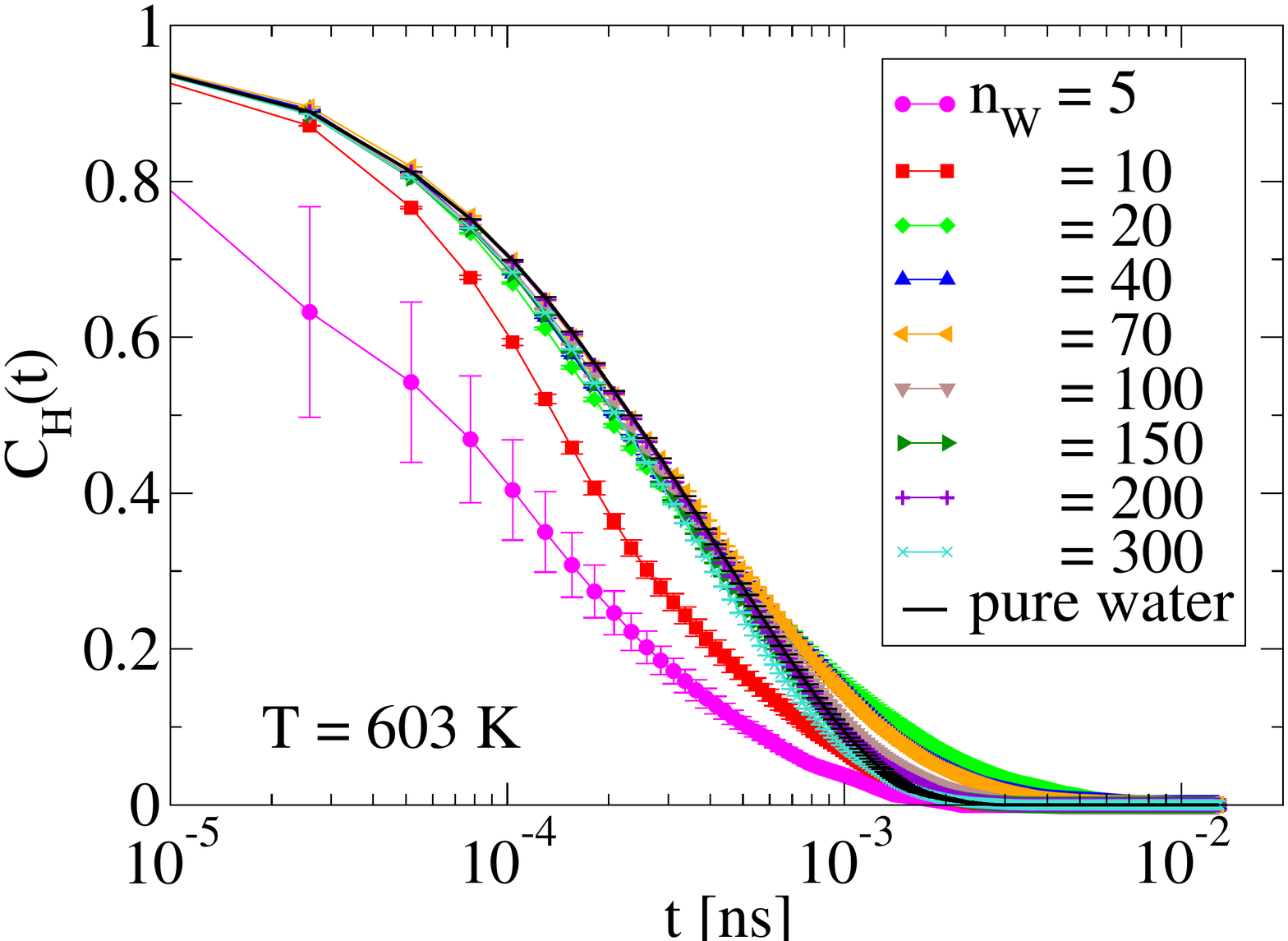}}
  \caption{\label{fig:corrHbond} Variation of the autocorrelation function
$C_H(t)$ between hydrogen bonds
with time for different number of water molecules and temperatures $T = 377 K$ (a),
$T = 452 K$ (b), and $T = 603 K$ (c).
}
  \end{center}
\end{figure}

\begin{figure}
  \begin{center}
  \includegraphics[scale=0.3]{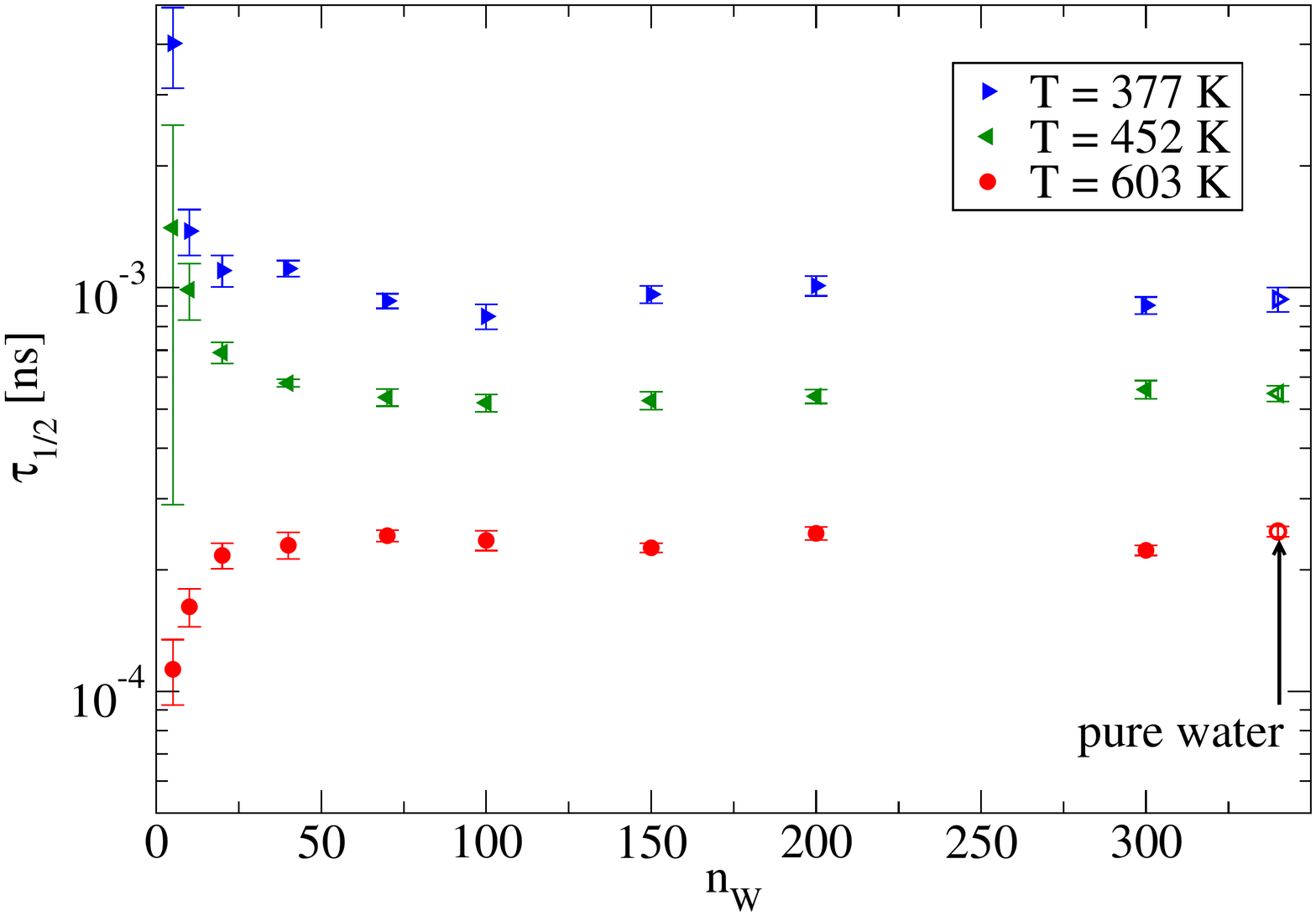}
  \caption{\label{fig:halfTime} Variation of the half lifetime of an hydrogen bond
$\tau_{1/2}$ with the number of water molecules $n_W$
for different temperatures. The open symbols correspond to the half lifetime in pure
water at the same temperatures.}
  \end{center}
\end{figure}

To supplement the description of water dynamics in bitumen at
short time scales, we looked at the hydrogen bond dynamics.
Hydrogen bonds can be defined using an energetic or a geometric criterion
in MD.~\cite{luzar93, marti93}
We choose the geometrical definition described in Ref.,~\cite{marti96}
adapted to the water model used in this work.
To be declared as bonded via a hydrogen bond, two water molecules should comply with the criteria:
\begin{enumerate}
\item The distance between the two oxygen atoms is less than $R_{OO} = 3.6$ \AA.
\item The distance between the hydrogen atom of one molecule and the oxygen atom
of the other is less than $R_{OH} = 2.45$ \AA.
\item The angle between the oxygen atom of one molecule, the oxygen atom,
and the hydrogen of atom the other molecule is less than $\phi = 30$\degree.
\end{enumerate}
We used the following definition for the autocorrelation function $C_H(t)$ between
hydrogen bonds:
\begin{equation}
C_H(t) = \frac{\langle h_{ij}(t_0, 0) h_{ij}(t_0, t)\rangle}{\langle h_{ij}(t_0, 0)\rangle},
\end{equation}
where $h_{ij}(t_0, t) = 1$ if the two water molecules $i$ and $j$ have been bonded without
any breaking between time $t_0$ and time $t$
and $h_{ij}(t_0, t) = 0$ otherwise. The average $\langle \cdot \rangle$
is done over all pairs $(i, j)$, with $i\neq j$, and over initial times $t_0$.
In this way, the correlation function is related to the lifetime of a single hydrogen bond.~\cite{marti96}
The time evolution of the correlation function $C_H(t)$ is plotted
in Figs.~\ref{fig:corrHbond} (a), (b), and (c) for different water contents
and at temperature $T = 377$ K, $T = 452$ K, and $T = 603$ K, respectively.
At temperature $T = 377$ K, the correlation functions associated
to water molecules in bitumen tend to decay more slowly than
the correlation function in bulk water. The more water molecules in bitumen,
the closer to the correlation function of bulk water.
It means that the droplet formed when the total number of water molecules
$n_W$ is low is stiffer than in bulk water.
The same result is shown in another form in Fig.~\ref{fig:halfTime},
displaying the variation of the half lifetime $\tau_{1/2}$ 
with the number of water molecules $n_W$ at different temperatures.
The half lifetime $\tau_{1/2}$ is defined as usual by $C_H(\tau_{1/2}) = 0.5$.
Figure~\ref{fig:halfTime} shows that the lifetime of a hydrogen bond
at the lowest temperature is around $1$ ps, which is in agreement
with other simulation~\cite{marti96} and experimental~\cite{woutersen} results.
At these time scales, according to Fig.~\ref{fig:msdW} (a),
the mean squared displacement of water molecules in the small stiff droplet is higher
than that of bulk water.
For a higher temperature $T = 452$ K a similar trend is observed:
the higher the total number of water molecules $n_W$, the smaller the half lifetime $\tau_{1/2}$.
At temperature $T = 603$ K, the inverse trend is observed.
To explain this, one can note that the cases of a total number of water molecules $n_W = 5$ 
at $T = 452$ K and $T = 603$ K, and $n_W = 10$ at $T = 603$ K
are singled out. The error bars on these results are quite large, because the initial
number of water molecules linked by a hydrogen bond is low at these high temperatures
and for such low numbers of water molecules. Indeed, Fig.~\ref{fig:drop_Vdistrib} shows that
the fraction of free water molecules in these 3 cases is very high.
This leads to poor statistics
on the correlation function and the overall results that these bonds do not last.

\section{Summary and Conclusions}
\label{sec:conclu}

In a system containing Cooee bitumen and up to $4$ \% in mass of water, 
molecular dynamics simulations have shown several features that describe
  how water molecules self-organize. 
(i) There is a loss of cohesion in the bitumen when
the water content is increased. But at temperature $T = 377$ K, this loss is due mainly
to the fact that the interactions with water molecules do not increase enough to
counteract the increase in volume.
At very high temperatures, the overall loss of cohesion is also due
to a loss of cohesion among the saturates molecules. The internal cohesion and structure of the aromatic
molecules, assembled into nanoaggregates, are unchanged as the water content increases.

(ii) The water molecules tend to form a droplet, which is mainly located next
  to saturates.  Most water is in a single large droplet at 452~K, while
  some smaller droplets occurred initially at 377~K.  A main droplet becomes dominant
  at 603~K only for the higher water concentrations studied. The number of
  water molecules outside the largest droplet reaches a plateau as the water
  concentration increases.
  The droplet volume increases linearly with the
  number of water molecules in the entire system.

(iii) The structure and dynamics of the water molecules are deeply affected
by the presence of bitumen.  At long times,
the water molecules reach a diffusive behavior, but it is governed
by the diffusion of the water droplet. The diffusion constant is consequently
much lower than that of bulk water and decreases as the droplet mass increases.
At short times, the hydrogen bond lifetime is higher in bitumen than in pure water
at temperature $T = 377$ K and for a small number of water molecules. The droplet is stiff.
When the number of water molecules increases, the hydrogen bond breaking of bulk water is recovered.
On the contrary, at very high temperatures $T = 603$ K, the hydrogen bond dynamics is 
faster in bitumen containing a small number of water molecules than in bulk water. This is due
to the fact that at these high temperatures, the water molecules need to be more numerous to form
a hydrogen bonded network.

Two natural perspectives arise from the conclusions just drawn. The first one
is to describe in more detail the dynamics of free water molecules. How long
do they spend outside the main droplet? How far can they go?
The second perspective is to study how the water structure and dynamics are affected
in the presence of hydrophilic surfaces such as those of filler particles and aggregates.
How many water molecules are left in bitumen in this case? How fast do the water
molecules reach the hydrophilic surfaces.


\providecommand*{\mcitethebibliography}{\thebibliography}
\csname @ifundefined\endcsname{endmcitethebibliography}
{\let\endmcitethebibliography\endthebibliography}{}


\begin{mcitethebibliography}{32}
\providecommand*{\natexlab}[1]{#1}
\providecommand*{\mciteSetBstSublistMode}[1]{}
\providecommand*{\mciteSetBstMaxWidthForm}[2]{}
\providecommand*{\mciteBstWouldAddEndPuncttrue}
  {\def\EndOfBibitem{\unskip.}}
\providecommand*{\mciteBstWouldAddEndPunctfalse}
  {\let\EndOfBibitem\relax}
\providecommand*{\mciteSetBstMidEndSepPunct}[3]{}
\providecommand*{\mciteSetBstSublistLabelBeginEnd}[3]{}
\providecommand*{\EndOfBibitem}{}
\mciteSetBstSublistMode{f}
\mciteSetBstMaxWidthForm{subitem}
{(\emph{\alph{mcitesubitemcount}})}
\mciteSetBstSublistLabelBeginEnd{\mcitemaxwidthsubitemform\space}
{\relax}{\relax}

\bibitem[Airey \emph{et~al.}(2008)Airey, Collop, Zoorob, and Elliott]{airey}
G.~D. Airey, A.~C. Collop, S.~E. Zoorob and R.~C. Elliott, \emph{Constr. Build.
  Mater.}, 2008, \textbf{22}, 2015--2024\relax
\mciteBstWouldAddEndPuncttrue
\mciteSetBstMidEndSepPunct{\mcitedefaultmidpunct}
{\mcitedefaultendpunct}{\mcitedefaultseppunct}\relax
\EndOfBibitem
\bibitem[Cui \emph{et~al.}(2014)Cui, Blackman, Kinloch, and Taylor]{cui}
S.~Cui, B.~R.~K. Blackman, A.~J. Kinloch and A.~C. Taylor, \emph{Int. J. Adhes.
  Adhes.}, 2014, \textbf{54}, 100--111\relax
\mciteBstWouldAddEndPuncttrue
\mciteSetBstMidEndSepPunct{\mcitedefaultmidpunct}
{\mcitedefaultendpunct}{\mcitedefaultseppunct}\relax
\EndOfBibitem
\bibitem[Blackman \emph{et~al.}(2013)Blackman, Cui, Kinloch, and
  Taylor]{blackman}
B.~R.~K. Blackman, S.~Cui, A.~J. Kinloch and A.~C. Taylor, \emph{Int. J. Adhes.
  Adhes.}, 2013, \textbf{42}, 1--10\relax
\mciteBstWouldAddEndPuncttrue
\mciteSetBstMidEndSepPunct{\mcitedefaultmidpunct}
{\mcitedefaultendpunct}{\mcitedefaultseppunct}\relax
\EndOfBibitem
\bibitem[Muthen(1998)]{muthen}
K.~M. Muthen, \emph{Foamed Asphalt Mixes - Mix Design Procedure}, External
  Contract Report CR-98/077, 1998\relax
\mciteBstWouldAddEndPuncttrue
\mciteSetBstMidEndSepPunct{\mcitedefaultmidpunct}
{\mcitedefaultendpunct}{\mcitedefaultseppunct}\relax
\EndOfBibitem
\bibitem[Salou \emph{et~al.}(1998)Salou, Siffert, and Jada]{salou}
M.~Salou, B.~Siffert and A.~Jada, \emph{Colloids Surf., A}, 1998, \textbf{142},
  9--16\relax
\mciteBstWouldAddEndPuncttrue
\mciteSetBstMidEndSepPunct{\mcitedefaultmidpunct}
{\mcitedefaultendpunct}{\mcitedefaultseppunct}\relax
\EndOfBibitem
\bibitem[Kasongo \emph{et~al.}(2000)Kasongo, Zhou, Xu, and Masliyah]{kasongo}
T.~Kasongo, Z.~Zhou, Z.~Xu and J.~Masliyah, \emph{Can. J. Chem. Eng.}, 2000,
  \textbf{78}, 674--681\relax
\mciteBstWouldAddEndPuncttrue
\mciteSetBstMidEndSepPunct{\mcitedefaultmidpunct}
{\mcitedefaultendpunct}{\mcitedefaultseppunct}\relax
\EndOfBibitem
\bibitem[Mullins(2011)]{mullins2011}
O.~C. Mullins, \emph{Annu. Rev. Anal. Chem.}, 2011, \textbf{4}, 393--418\relax
\mciteBstWouldAddEndPuncttrue
\mciteSetBstMidEndSepPunct{\mcitedefaultmidpunct}
{\mcitedefaultendpunct}{\mcitedefaultseppunct}\relax
\EndOfBibitem
\bibitem[Gray \emph{et~al.}(2011)Gray, Tykwinski, Stryker, and Tan]{gray}
M.~R. Gray, R.~R. Tykwinski, J.~M. Stryker and X.~Tan, \emph{Energy \& Fuels},
  2011, \textbf{25}, 3125--3134\relax
\mciteBstWouldAddEndPuncttrue
\mciteSetBstMidEndSepPunct{\mcitedefaultmidpunct}
{\mcitedefaultendpunct}{\mcitedefaultseppunct}\relax
\EndOfBibitem
\bibitem[Hansen \emph{et~al.}(2013)Hansen, Lemarchand, Nielsen, and
  Dyre]{hansen_jcp_2013}
J.~S. Hansen, C.~A. Lemarchand, E.~Nielsen and J.~C. Dyre, \emph{J. Chem.
  Phys.}, 2013, \textbf{138}, 094508\relax
\mciteBstWouldAddEndPuncttrue
\mciteSetBstMidEndSepPunct{\mcitedefaultmidpunct}
{\mcitedefaultendpunct}{\mcitedefaultseppunct}\relax
\EndOfBibitem
\bibitem[Hubbard and Stanfield(1948)]{hubbard:1948}
R.~Hubbard and K.~Stanfield, \emph{Anal. Chem.}, 1948, \textbf{20}, 460\relax
\mciteBstWouldAddEndPuncttrue
\mciteSetBstMidEndSepPunct{\mcitedefaultmidpunct}
{\mcitedefaultendpunct}{\mcitedefaultseppunct}\relax
\EndOfBibitem
\bibitem[{Jones IV}(1993)]{shrp645}
D.~R. {Jones IV}, \emph{Asphalt Cements: A Concise Data Compilation},
  SHRP-A-645, Strategic Highway Research Program, 1993\relax
\mciteBstWouldAddEndPuncttrue
\mciteSetBstMidEndSepPunct{\mcitedefaultmidpunct}
{\mcitedefaultendpunct}{\mcitedefaultseppunct}\relax
\EndOfBibitem
\bibitem[Rostler(1965)]{rostler:1965}
F.~Rostler, \emph{Asphalts, Vol. II}, John Wiley and {S}ons, 1965\relax
\mciteBstWouldAddEndPuncttrue
\mciteSetBstMidEndSepPunct{\mcitedefaultmidpunct}
{\mcitedefaultendpunct}{\mcitedefaultseppunct}\relax
\EndOfBibitem
\bibitem[Liu \emph{et~al.}(2010)Liu, Zhou, Zhang, and Zhang]{liu:2010}
X.~Liu, G.~Zhou, X.~Zhang and S.~Zhang, \emph{AIChE J.}, 2010, \textbf{56},
  2983\relax
\mciteBstWouldAddEndPuncttrue
\mciteSetBstMidEndSepPunct{\mcitedefaultmidpunct}
{\mcitedefaultendpunct}{\mcitedefaultseppunct}\relax
\EndOfBibitem
\bibitem[Allen and Tildesley(1987)]{AllenTildesley}
M.~P. Allen and D.~J. Tildesley, \emph{Computer Simulation of Liquids}, Oxford
  Science Publications, Oxford, 1987\relax
\mciteBstWouldAddEndPuncttrue
\mciteSetBstMidEndSepPunct{\mcitedefaultmidpunct}
{\mcitedefaultendpunct}{\mcitedefaultseppunct}\relax
\EndOfBibitem
\bibitem[Hansen \emph{et~al.}(2012)Hansen, der, and Dyre]{hansen_2012}
J.~S. Hansen, T.~B.~S. der and J.~C. Dyre, \emph{J. Phys. Chem. B}, 2012,
  \textbf{116}, 5738--5743\relax
\mciteBstWouldAddEndPuncttrue
\mciteSetBstMidEndSepPunct{\mcitedefaultmidpunct}
{\mcitedefaultendpunct}{\mcitedefaultseppunct}\relax
\EndOfBibitem
\bibitem[Takahashi \emph{et~al.}(2011)Takahashi, Narumi, and
  Yasouko]{takahashi_2011}
K.~Takahashi, T.~Narumi and K.~Yasouko, \emph{J. Chem. Phys.}, 2011,
  \textbf{134}, 174112\relax
\mciteBstWouldAddEndPuncttrue
\mciteSetBstMidEndSepPunct{\mcitedefaultmidpunct}
{\mcitedefaultendpunct}{\mcitedefaultseppunct}\relax
\EndOfBibitem
\bibitem[Toukan and Rahman(1985)]{toukan_1985}
K.~Toukan and A.~Rahman, \emph{Phys. Rev. B}, 1985, \textbf{31}, 2643\relax
\mciteBstWouldAddEndPuncttrue
\mciteSetBstMidEndSepPunct{\mcitedefaultmidpunct}
{\mcitedefaultendpunct}{\mcitedefaultseppunct}\relax
\EndOfBibitem
\bibitem[Wu \emph{et~al.}(2006)Wu, Tepper, and Voth]{wu_2006}
Y.~Wu, H.~J. Tepper and G.~A. Voth, \emph{J. Chem. Phys.}, 2006, \textbf{124},
  024503\relax
\mciteBstWouldAddEndPuncttrue
\mciteSetBstMidEndSepPunct{\mcitedefaultmidpunct}
{\mcitedefaultendpunct}{\mcitedefaultseppunct}\relax
\EndOfBibitem
\bibitem[Raabe and Sadus(2007)]{raabe_2007}
G.~Raabe and R.~J. Sadus, \emph{J. Chem. Phys.}, 2007, \textbf{126},
  044701\relax
\mciteBstWouldAddEndPuncttrue
\mciteSetBstMidEndSepPunct{\mcitedefaultmidpunct}
{\mcitedefaultendpunct}{\mcitedefaultseppunct}\relax
\EndOfBibitem
\bibitem[Bailey \emph{et~al.}(2015)Bailey, Ingebrigtsen, Hansen, Veldhorst,
  B{\'o}hling, Lemarchand, Olsen, Bacher, Larsen, Dyre, and Schr{\'o}der]{rumd}
N.~Bailey, T.~Ingebrigtsen, J.~Hansen, A.~Veldhorst, L.~B{\'o}hling,
  C.~Lemarchand, A.~Olsen, A.~Bacher, H.~Larsen, J.~Dyre and T.~Schr{\'o}der,
  \emph{submitted to J. Comput. Chem.}, 2015\relax
\mciteBstWouldAddEndPuncttrue
\mciteSetBstMidEndSepPunct{\mcitedefaultmidpunct}
{\mcitedefaultendpunct}{\mcitedefaultseppunct}\relax
\EndOfBibitem
\bibitem[Greenfield and Theodorou(1993)]{greenfield93}
M.~L. Greenfield and D.~N. Theodorou, \emph{Macromolecules}, 1993, \textbf{26},
  5461--5472\relax
\mciteBstWouldAddEndPuncttrue
\mciteSetBstMidEndSepPunct{\mcitedefaultmidpunct}
{\mcitedefaultendpunct}{\mcitedefaultseppunct}\relax
\EndOfBibitem
\bibitem[Tanemura \emph{et~al.}(1983)Tanemura, Ogawa, and Ogita]{tanemura83}
M.~Tanemura, T.~Ogawa and N.~Ogita, \emph{J. Comput. Phys.}, 1983, \textbf{51},
  191--207\relax
\mciteBstWouldAddEndPuncttrue
\mciteSetBstMidEndSepPunct{\mcitedefaultmidpunct}
{\mcitedefaultendpunct}{\mcitedefaultseppunct}\relax
\EndOfBibitem
\bibitem[Dack(1975)]{dack}
M.~R.~J. Dack, \emph{Chem. Soc. Rev.}, 1975, \textbf{4}, 211--229\relax
\mciteBstWouldAddEndPuncttrue
\mciteSetBstMidEndSepPunct{\mcitedefaultmidpunct}
{\mcitedefaultendpunct}{\mcitedefaultseppunct}\relax
\EndOfBibitem
\bibitem[Redelius(2000)]{redelius}
P.~G. Redelius, \emph{Fuel}, 2000, \textbf{79}, 27\relax
\mciteBstWouldAddEndPuncttrue
\mciteSetBstMidEndSepPunct{\mcitedefaultmidpunct}
{\mcitedefaultendpunct}{\mcitedefaultseppunct}\relax
\EndOfBibitem
\bibitem[Barton(1974)]{barton}
A.~F.~M. Barton, \emph{Chem. Rev.}, 1974, \textbf{75}, 731\relax
\mciteBstWouldAddEndPuncttrue
\mciteSetBstMidEndSepPunct{\mcitedefaultmidpunct}
{\mcitedefaultendpunct}{\mcitedefaultseppunct}\relax
\EndOfBibitem
\bibitem[Mullins \emph{et~al.}(2012)Mullins, Sabbah, Eyssautier, Pomerantz,
  Barr\'e, Andrews, Ruiz-Morales, Mostowfi, McFarlane, Goual, Lepkowicz,
  Cooper, Orbulescu, Leblanc, Edwards, and Zare]{mullins2012}
O.~C. Mullins, H.~Sabbah, J.~Eyssautier, A.~E. Pomerantz, L.~Barr\'e, A.~B.
  Andrews, Y.~Ruiz-Morales, F.~Mostowfi, R.~McFarlane, L.~Goual, R.~Lepkowicz,
  T.~Cooper, J.~Orbulescu, R.~M. Leblanc, J.~Edwards and R.~N. Zare,
  \emph{Energy \& Fuels}, 2012, \textbf{26}, 3986--4003\relax
\mciteBstWouldAddEndPuncttrue
\mciteSetBstMidEndSepPunct{\mcitedefaultmidpunct}
{\mcitedefaultendpunct}{\mcitedefaultseppunct}\relax
\EndOfBibitem
\bibitem[Lemarchand \emph{et~al.}(2013)Lemarchand, der, Dyre, and
  Hansen]{aging}
C.~A. Lemarchand, T.~B.~S. der, J.~C. Dyre and J.~S. Hansen, \emph{J. Chem.
  Phys.}, 2013, \textbf{138}, 094508\relax
\mciteBstWouldAddEndPuncttrue
\mciteSetBstMidEndSepPunct{\mcitedefaultmidpunct}
{\mcitedefaultendpunct}{\mcitedefaultseppunct}\relax
\EndOfBibitem
\bibitem[Lemarchand and Hansen(2015)]{branched}
C.~A. Lemarchand and J.~S. Hansen, \emph{accepted in J. Phys. Chem. B},
  2015\relax
\mciteBstWouldAddEndPuncttrue
\mciteSetBstMidEndSepPunct{\mcitedefaultmidpunct}
{\mcitedefaultendpunct}{\mcitedefaultseppunct}\relax
\EndOfBibitem
\bibitem[Luzar and Chandler(1993)]{luzar93}
A.~Luzar and D.~Chandler, \emph{J. Chem. Phys.}, 1993, \textbf{98}, 8160\relax
\mciteBstWouldAddEndPuncttrue
\mciteSetBstMidEndSepPunct{\mcitedefaultmidpunct}
{\mcitedefaultendpunct}{\mcitedefaultseppunct}\relax
\EndOfBibitem
\bibitem[Padro \emph{et~al.}(1994)Padro, Mart\'i, and Gu\'ardia]{marti93}
J.~A. Padro, J.~Mart\'i and E.~Gu\'ardia, \emph{J. Phys.: Condens. Matter},
  1994, \textbf{6}, 2283--2290\relax
\mciteBstWouldAddEndPuncttrue
\mciteSetBstMidEndSepPunct{\mcitedefaultmidpunct}
{\mcitedefaultendpunct}{\mcitedefaultseppunct}\relax
\EndOfBibitem
\bibitem[Mart\'i \emph{et~al.}(1996)Mart\'i, Padro, and Gu\'ardia]{marti96}
J.~Mart\'i, J.~A. Padro and E.~Gu\'ardia, \emph{J. Chem. Phys.}, 1996,
  \textbf{105}, 639\relax
\mciteBstWouldAddEndPuncttrue
\mciteSetBstMidEndSepPunct{\mcitedefaultmidpunct}
{\mcitedefaultendpunct}{\mcitedefaultseppunct}\relax
\EndOfBibitem
\bibitem[Woutersen \emph{et~al.}(1998)Woutersen, Emmerichs, Nienhuys, and
  Bakker]{woutersen}
S.~Woutersen, U.~Emmerichs, H.-K. Nienhuys and H.~J. Bakker, \emph{Phys. Rev.
  Lett.}, 1998, \textbf{81}, 1106\relax
\mciteBstWouldAddEndPuncttrue
\mciteSetBstMidEndSepPunct{\mcitedefaultmidpunct}
{\mcitedefaultendpunct}{\mcitedefaultseppunct}\relax
\EndOfBibitem
\end{mcitethebibliography}

\end{document}